\documentclass[12pt]{article}
\usepackage{amssymb}
\usepackage{rotating}
\usepackage{amsmath}
\usepackage{arydshln}
\allowdisplaybreaks[4]

\renewcommand{\theequation}{\arabic{section}.\arabic{equation}}

\begin{document}



\def\a{\alpha}
\def\b{\beta}
\def\d{\delta}
\def\e{\epsilon}
\def\g{\gamma}
\def\h{\mathfrak{h}}
\def\k{\kappa}
\def\l{\lambda}
\def\o{\omega}
\def\p{\wp}
\def\r{\rho}
\def\t{\tau}
\def\s{\sigma}
\def\z{\zeta}
\def\x{\xi}
\def\V={{{\bf\rm{V}}}}
 \def\A{{\cal{A}}}
 \def\B{{\cal{B}}}
 \def\C{{\cal{C}}}
 \def\D{{\cal{D}}}
\def\G{\Gamma}
\def\K{{\cal{K}}}
\def\O{\Omega}
\def\R{\bar{R}}
\def\T{{\cal{T}}}
\def\L{\Lambda}
\def\f{E_{\tau,\eta}(sl_2)}
\def\E{E_{\tau,\eta}(sl_n)}
\def\Zb{\mathbb{Z}}
\def\Cb{\mathbb{C}}

\def\R{\overline{R}}

\def\beq{\begin{equation}}
\def\eeq{\end{equation}}
\def\bea{\begin{eqnarray}}
\def\eea{\end{eqnarray}}
\def\ba{\begin{array}}
\def\ea{\end{array}}
\def\no{\nonumber}
\def\le{\langle}
\def\re{\rangle}
\def\lt{\left}
\def\rt{\right}

\newtheorem{Theorem}{Theorem}
\newtheorem{Definition}{Definition}
\newtheorem{Proposition}{Proposition}
\newtheorem{Lemma}{Lemma}
\newtheorem{Corollary}{Corollary}
\newcommand{\proof}[1]{{\bf Proof. }
        #1\begin{flushright}$\Box$\end{flushright}}

\baselineskip=20pt

\newfont{\elevenmib}{cmmib10 scaled\magstep1}
\newcommand{\preprint}{
   \begin{flushleft}
   \end{flushleft}\vspace{-1.3cm}
   \begin{flushright}\normalsize
   \end{flushright}}
\newcommand{\Title}[1]{{\baselineskip=26pt
   \begin{center} \Large \bf #1 \\ \ \\ \end{center}}}
\newcommand{\Author}{\begin{center}
   \large \bf
Guang-Liang Li${}^{a}$, Junpeng Cao${}^{b,c,d}$, Panpan
Xue${}^{a}$,  Kun Hao${}^{e,f}$, Pei Sun${}^{e,f}$, Wen-Li
Yang${}^{e,f,g}\footnote{Corresponding author:
wlyang@nwu.edu.cn}$, Kangjie Shi${}^{e,f}$ and Yupeng
Wang${}^{b,c,h}\footnote{Corresponding author: yupeng@iphy.ac.cn}$
 \end{center}}

\newcommand{\Address}{\begin{center}

     ${}^a$ Department of Applied Physics, Xi'an Jiaotong University, Xian 710049, China\\
     ${}^b$ Beijing National Laboratory for Condensed Matter
           Physics, Institute of Physics, Chinese Academy of Sciences, Beijing
           100190, China\\
     ${}^c$ School of Physical Sciences, University of Chinese Academy of
Sciences, Beijing, China\\
     ${}^d$ Songshan Lake Materials Laboratory, Dongguan, Guangdong 523808, China \\
     ${}^e$ Institute of Modern Physics, Northwest University,
     Xian 710127, China\\
     ${}^f$ Shaanxi Key Laboratory for Theoretical Physics Frontiers,  Xian 710127, China\\
     ${}^g$ School of Physics, Northwest University,  Xian 710127, China\\
     ${}^h$ The Yangtze River Delta Physics Research Center, Liyang, Jiangsu, China

   \end{center}}
\newcommand{\Accepted}[1]{\begin{center}
   {\large \sf #1}\\ \vspace{1mm}{\small \sf Accepted for Publication}
   \end{center}}

\preprint
\thispagestyle{empty}
\bigskip\bigskip\bigskip

\Title{Off-diagonal Bethe Ansatz for the $D^{(1)}_3$ model}

\Author

\Address
\vspace{1cm}

\begin{abstract}

The exact solutions of the $D^{(1)}_3$ model (or the $so(6)$  quantum spin chain) with either periodic or general integrable open boundary conditions are obtained by using the off-diagonal Bethe Ansatz. From the fusion, the complete operator product identities are obtained, which are sufficient  to enable us to  determine  spectrum of the system.
Eigenvalues of the fused transfer matrices are constructed by the $T-Q$ relations for the periodic case and by the inhomogeneous $T-Q$ one for the non-diagonal
boundary reflection case. The present method can be generalized to deal with the $D^{(1)}_{n}$ model directly.

\vspace{1truecm} \noindent {\it PACS:} 75.10.Pq, 02.30.Ik, 71.10.Pm

\end{abstract}
\newpage
\section{Introduction}
\label{intro} \setcounter{equation}{0}

Quantum integrable models play an important role in the condensed matter, cold atoms, theoretical and mathematical physics \cite{1,Kor93}.
Recently, the models with general integrable non-diagonal boundary reflections attract many attentions. Due to the $U(1)$-symmetry broken,
the traditional Bethe Ansatz does not work. People developed many interesting methods, such as q-Onsager algebra \cite{q-1,q-2,q-3,q-4}, separation of variables \cite{Fra08,Fra11,Nic12},
modified algebraic Bethe Ansatz \cite{Bel13,Bel15,Bel15-1,Ava15}, off-diagonal Bethe Ansatz (ODBA) \cite{Cao1,Cao2} and others \cite{Nep13}.

The exact solution of integrable models associated with high rank Lie algebra is a very interesting issue. The case that the models with periodic or diagonal boundary reflection, where the $U(1)$-charge is conserved,
has been studied extensively. For example, Reshetikhin derived the energy spectrum of the periodic quantum integrable models associated with $B_n$, $C_n$, $D_n$ and other Lie algebras by using the analytic Bethe Ansatz \cite{NYReshetikhin1,NYReshetikhin2}. Martins and Ramos studied this kind of models with periodic boundary condition by using the algebraic Bethe Ansatz \cite{Bn}. Li, Shi and Yue investigated the
open boundary cases where the reflection matrices only have the diagonal elements \cite{b2aba,c2aba}.

The ODBA is a universal method to treat the quantum integrable models with or without $U(1)$-symmetry, even for the high rank cases.
The nested ODBA was proposed to study the spin chain associated with $A_n$ Lie algebra with generic non-diagonal boundary reflections \cite{Cao14JHEP143,Cao15JHEP036,Hao16}.
The ODBA also has been applied to the models associated with $A_2^{(2)}$ \cite{Hao14}, $B_2$ \cite{Li19-1} and $C_2$ \cite{Li19} Lie algebras.

In this paper, we develop a nested ODBA method to approach the quantum integrable $D^{(1)}_3$ (simplest case of $D^{(1)}_n$) model with either periodic or non-diagonal open boundary conditions.
We have obtained the complete operator product identities based on the fusion \cite{Kar79-1,Kar79-10,Kar79-2,Kar79-3} and the eigenvalues of the system based on the (inhomogeneous) $T-Q$ relations.

The paper is organized as follows. In section 2, we review the $R$-matrix associated with the  $D^{(1)}_3$ model, which is the starting point.
In section 3, we
construct the $D^{(1)}_3$ model with periodic boundary condition. The Hamiltonian,
closed functional relations among the eigenvalues of the fused transfer matrices, and the spectrum of the system are obtained with fusion techniques.
In section 4, we study the $D^{(1)}_3$ model with an off-diagonal open boundary reflection.
By constructing some operator product identities, we derive the exact eigenvalues of the transfer matrix in terms of an inhomogeneous $T-Q$ relation. Section 5 is attributed to concluding remarks.


\section{The $R$-matrix}
\setcounter{equation}{0}

Throughout, ${\rm\bf V}$ denotes a six-dimensional linear space  which endows the fundamental vectorial representation of  $so(6)$ (or the $D_3$) algebra, and let $\{|i\rangle, i=1,2,\cdots,6\}$ be an orthogonal basis of it.
We adopt the standard notations: for any matrix $A\in {\rm End}({\rm\bf V})$, $A_j$ is an
embedding operator in the tensor space ${\rm\bf V}\otimes
{\rm\bf V}\otimes\cdots$, which acts as $A$ on the $j$-th space and as
identity on the other factor spaces; For $B\in {\rm End}({\rm\bf V}\otimes {\rm\bf V})$, $B_{ij}$ is an embedding
operator of $B$ in the tensor space, which acts as identity
on the factor spaces except for the $i$-th and $j$-th ones.

The $R$-matrix $R(u)\in {\rm End}({\rm\bf V}\otimes {\rm\bf V})$
of the $D^{(1)}_3$ model is
\begingroup
\renewcommand*{\arraystretch}{0.1}
\begin{eqnarray}
R^{vv}_{ij}= \begin{pmatrix}\setlength{\arraycolsep}{2.0pt}
    \begin{array}{cccccc|cccccc|cccccc|cccccc|cccccc|cccccc}
    a&&&&& &&&&&& &&&&&& &&&&&& &&&&&& &&&&&& \\
    &b&&&& &g&&&&& &&&&&& &&&&&& &&&&&& &&&&&&\\
    &&b&&& &&&&&& &g&&&&& &&&&&& &&&&&& &&&&&& \\
    &&&b&& &&&&&& &&&&&& &g&&&&& &&&&&& &&&&&& \\
    &&&&b& &&&&&& &&&&&& &&&&&& &g&&&&& &&&&&& \\
    &&&&&e &&&&&d& &&&&d&& &&&d&&& &&d&&&& &f&&&&&\\
   \hline &g&&&& &b&&&&& &&&&&& &&&&&& &&&&&& &&&&&& \\
    &&&&& &&a&&&& &&&&&& &&&&&& &&&&&& &&&&&& \\
    &&&&& &&&b&&& &&g&&&& &&&&&& &&&&&& &&&&&& \\
    &&&&& &&&&b&& &&&&&& &&g&&&& &&&&&& &&&&&& \\
     &&&&&d &&&&&e& &&&&d&& &&&d&&& &&f&&&& &d&&&&&\\
     &&&&& &&&&&&b &&&&&& &&&&&& &&&&&& &&g&&&& \\
   \hline  &&g&&& &&&&&& &b&&&&& &&&&&& &&&&&& &&&&&& \\
    &&&&& &&&g&&& &&b&&&& &&&&&& &&&&&& &&&&&& \\
   &&&&& &&&&&& &&&a&&& &&&&&& &&&&&& &&&&&& \\
   &&&&&d &&&&&d& &&&&e&& &&&f&&& &&d&&&& &d&&&&&\\
     &&&&& &&&&&& &&&&&b& &&&&&& &&&g&&& &&&&&& \\
      &&&&& &&&&&& &&&&&&b &&&&&& &&&&&& &&&g&&& \\
   \hline &&&g&& &&&&&& &&&&&& &b&&&&& &&&&&& &&&&&& \\
    &&&&& &&&&g&& &&&&&& &&b&&&& &&&&&& &&&&&& \\
    &&&&&d  &&&&&d& &&&&f&& &&&e&&& &&d&&&& &d&&&&&\\
     &&&&& &&&&&& &&&&&& &&&&a&& &&&&&& &&&&&& \\
     &&&&& &&&&&& &&&&&& &&&&&b& &&&&g&& &&&&&& \\
      &&&&& &&&&&& &&&&&& &&&&&&b &&&&&& &&&&g&& \\
    \hline  &&&&g& &&&&&& &&&&&& &&&&&& &b&&&&& &&&&&& \\
    &&&&&d &&&&&f& &&&&d&& &&&d&&& &&e&&&& &d&&&&&\\
     &&&&& &&&&&& &&&&&g& &&&&&& &&&b&&& &&&&&& \\
    &&&&& &&&&&& &&&&&& &&&&&g& &&&&b&& &&&&&& \\
     &&&&& &&&&&& &&&&&& &&&&&& &&&&&a& &&&&&& \\
      &&&&& &&&&&& &&&&&& &&&&&& &&&&&&b &&&&&g& \\
    \hline &&&&&f &&&&&d& &&&&d&& &&&d&&& &&d&&&& &e&&&&&\\
    &&&&& &&&&&&g &&&&&& &&&&&& &&&&&& &&b&&&& \\
    &&&&& &&&&&& &&&&&&g &&&&&& &&&&&& &&&b&&& \\
     &&&&& &&&&&& &&&&&& &&&&&&g &&&&&& &&&&b&& \\
    &&&&& &&&&&& &&&&&& &&&&&& &&&&&&g  &&&&&b& \\
     &&&&& &&&&&& &&&&&& &&&&&& &&&&&& &&&&&&a \\
\end{array}
    \end{pmatrix}, \label{0804-1}
\end{eqnarray}
\endgroup where the matrix elements are \bea
&&a=(u+1)(u+2), \quad
b=u(u+2), \quad f=2,\nonumber\\[4pt]
&& d=-u,\quad e=u(u+1),\quad g=u+2, \no \eea
and $u$ is the spectral parameter.

The $R$-matrix (\ref{0804-1}) satisfies the properties
\begin{eqnarray}
\hspace{-0.8truecm}{\rm regularity}&:&R^{  vv}_{12}(0)=\rho_1(0)^{\frac{1}{2}}{\cal P}_{12},\nonumber\\[4pt]
\hspace{-0.8truecm}{\rm unitarity}&:&R^{   vv}_{12}(u)R^{  vv}_{21}(-u)=\rho_1(u)=a(u)a(-u),\nonumber\\[4pt]
\hspace{-0.8truecm}{\rm crossing \,symmetry}&:&R^{
vv}_{12}(u)=V_1\,\{R^{ vv}_{12}(-u-{2})\}^{t_2}\,V_1=V_2\,\{R^{
vv}_{12}(-u-{2})\}^{t_1}\,V_2,\label{Crossing-symmetry}
\end{eqnarray}
where ${\cal P}_{12}$ is the permutation operator with the matrix
elements $[{\cal P}_{12}]^{ij}_{kl}=\delta_{il}\delta_{jk}$, $t_i$
denotes the transposition in the $i$-th space, and $R _{21}={\cal
P}_{12}R _{12}{\cal P}_{12}$,
and the crossing-matrix $V$ is
\begin{eqnarray}
V=\lt(\begin{array}{cccccc}&&&&&1\\
&&&&1&\\&&&1&&\\&&1&&&\\&1&&&&\\
1&&&&&
\end{array}\rt),\quad V^2={\rm id}.
\end{eqnarray}
Combining  the crossing-symmetry and the unitarity of the $R$-matrix, one can derive the crossing-unitarity relation
\bea
{\rm crossing \,unitarity}&:&R^{
vv}_{12}(u)^{t_1}R^{
vv}_{21}(-u-4)^{t_1}=\rho_1(u+2).\label{Crossing-Unitarity}
\eea
The $R$-matrix (\ref{0804-1}) satisfies the Yang-Baxter equation
\begin{eqnarray}
R^{   vv}_{12}(u-v)R^{  vv}_{13}(u)R^{    vv}_{23}(v)=R^{
  vv}_{23}(v)R^{   vv}_{13}(u)R^{  vv}_{12}(u-v). \label{20190802-1}
\end{eqnarray}

\section{$D^{(1)}_3$ model with periodic boundary condition}

\setcounter{equation}{0}

For the periodic boundary condition, we introduce the monodromy matrix
\bea
T_0^{  v}(u)=R^{    vv}_{01}(u-\theta_1)R^{
vv}_{02}(u-\theta_2)\cdots R^{
   vv}_{0N}(u-\theta_N), \label{Mon-1}
\eea
where the index $0$ indicates the auxiliary space and the other tensor space  ${\rm\bf V}^{\otimes N}$ is the physical or quantum space, $N$ is the number of sites and $\{\theta_j\}$ are the inhomogeneous parameters.
The monodromy matrix satisfies the Yang-Baxter relation
\bea
 R^{  vv}_{12}(u-v) T_1^{  v}(u) T_2^{  v}(v) = T_2^{  v}(v) T_1^{  v}(u) R^{  vv}_{12}(u-v).\label{ybta2o}
\eea
The transfer matrix is given by the trace of monodromy matrix in the auxiliary space
\bea t^{(p)}(u)=tr_0 T_0^{  v}(u). \label{1117-1}\eea
From the Yang-Baxter relation (\ref{ybta2o}), one can prove that the transfer matrices with different spectral parameters
commute with each other, $[t^{(p)}(u), t^{(p)}(v)]=0$. Therefore, $t^{(p)}(u)$ serves
as the generating function of the conserved quantities of the
system. The Hamiltonian is given  in terms of  the transfer matrix (\ref{1117-1})  as
\begin{eqnarray}
H_p= \frac{\partial \ln t^{(p)}(u)}{\partial
u}|_{u=0,\{\theta_j\}=0}=\sum_{k=1}^{N}H_{k\,k+1},\label{haha-1}
\end{eqnarray}
where
\bea
H_{k\,k+1}= {\cal P}_{k\,k+1}\,\frac{\partial}{\partial u}R_{k\,k+1}(u)\left.\right|_{u=0}, \label{local-Ham}
\eea
and the periodic boundary condition reads
\bea
H_{N\,N+1}=H_{N\,1}.\label{Periodic-boundary-condition}
\eea

\subsection{The fusion}

At some special points, the $R$-matrix (\ref{0804-1}) degenerates into the projector operators which enables us to do the fusion.
For example, at the point of $u=-2$, we have
\bea R^{vv}_{12}(-2)=P^{{  vv}(1) }_{12}S_{12}^{(1)}, \eea
where $S_{12}^{(1)}$ is a constant matrix $\in {\rm End}({\rm\bf V}\otimes {\rm\bf V})$ omitted here, $P^{{  vv}(1) }_{12}$ is the one-dimensional projector \bea P^{{
vv}(1) }_{12}=|\psi_0\rangle\langle\psi_0|, \quad  P^{{ vv}(1)
}_{21}= P^{{  vv}(1) }_{12}, \label{a1} \eea and the basis vector reads
\bea
|\psi_0\rangle=\frac{1}{\sqrt{6}}(|16\rangle+|25\rangle+|34\rangle+|43\rangle+|52\rangle+|61\rangle).\no
\eea
From the Yang-Baxter equation (\ref{20190802-1}), we obtain the following fusion identities
\bea
&&P^{ {  vv}(1) }_{21}R^{vv} _{13}(u)R^{vv} _{23}(u-2)P^{ {  vv}(1) }_{21}=a(u)e(u-2)P^{{  vv} (1) }_{21},\nonumber\\
\label{fu-1} &&P^{ {  vv}(1) }_{12}R^{vv} _{31}(u)R^{vv}
_{32}(u-2)P^{ { vv}(1) }_{12}=a(u)e(u-2)P^{ {  vv}(1) }_{12}.
\label{hhgg-1} \eea

At the point of $u=-1$, we have \bea R^{vv}_{12}(-1)=P^{{  vv}
(16) }_{12}S_{12}^{(16)},\eea where $S_{12}^{(16)}$ is a constant
matrix omitted here and $P^{{  vv} (16) }_{12}$ is a
16-dimensional projector \bea P^{{  vv} (16)
}_{12}=\sum_{i=1}^{16}
|{\phi}^{(16)}_i\rangle\langle{\phi}^{(16)}_i|,\quad P^{{  vv}
(16) }_{21} =P^{{  vv} (16)
}_{12}|_{1\rightarrow2,2\rightarrow1},\no \eea with the basis
vectors:
 \bea
&&|{\phi}^{(16)}_1\rangle=\frac{1}{\sqrt{2}}(|12\rangle-|21\rangle),
\quad |{\phi}^{(16)}_2\rangle=\frac{1}{\sqrt{2}}(|13\rangle-|31\rangle),\quad
|{\phi}^{(16)}_3\rangle=\frac{1}{\sqrt{2}}(|14\rangle-|41\rangle), \nonumber\\[4pt]
&& |{\phi}^{(16)}_4\rangle=\frac{1}{\sqrt{2}}(|15\rangle-|51\rangle),\quad
|{\phi}^{(16)}_5\rangle=\frac{1}{\sqrt{2}}(|16\rangle-|61\rangle), \quad
|{\phi}^{(16)}_6\rangle=\frac{1}{\sqrt{2}}(|23\rangle-|32\rangle), \nonumber\\[4pt]
&& |{\phi}^{(16)}_7\rangle=\frac{1}{\sqrt{2}}(|24\rangle-|42\rangle),\quad
|{\phi}^{(16)}_8\rangle=\frac{1}{\sqrt{2}}(|25\rangle-|52\rangle),\quad
|{\phi}^{(16)}_9\rangle=\frac{1}{\sqrt{2}}(|26\rangle-|62\rangle),\nonumber\\[4pt]
&& |{\phi}^{(16)}_{10}\rangle=\frac{1}{\sqrt{2}}(|34\rangle-|43\rangle),\quad
|{\phi}^{(16)}_{11}\rangle=\frac{1}{\sqrt{2}}(|35\rangle-|53\rangle),\quad |{\phi}^{(16)}_{12}\rangle=\frac{1}{\sqrt{2}}(|36\rangle-|63\rangle),\nonumber\\[4pt]
&&|{\phi}^{(16)}_{13}\rangle=\frac{1}{\sqrt{6}}(|34\rangle+|43\rangle+|25\rangle+|52\rangle+|16\rangle+|61\rangle),\quad
|{\phi}^{(16)}_{14}\rangle=\frac{1}{\sqrt{2}}(|45\rangle-|54\rangle),\nonumber\\[4pt]
&&|{\phi}^{(16)}_{15}\rangle=\frac{1}{\sqrt{2}}(|46\rangle-|64\rangle),\quad
|{\phi}^{(16)}_{16}\rangle=\frac{1}{\sqrt{2}}(|56\rangle-|65\rangle).\no
\eea From the Yang-Baxter equation (\ref{20190802-1}) and the
16-dimensional projector $P^{{  vv} (16) }_{12}$, we obtain the
fusion identities \bea &&P^{{  vv} (16) }_{21}R^{vv}
_{13}(u)R^{vv} _{23}(u-1)P_{21}^{{  vv} (16)
}=(u-1)(u+2)S_{1'2'}R^{ s_+v}_{1'3}(u-\frac12)R^{
s_-v}_{2'3}(u-\frac12)S_{1'2'}^{-1},
\no \\
  &&P^{{  vv} (16) }_{12}R^{vv} _{31}(u)R^{vv}_{32}(u-1)P_{12}^{{  vv} (16) }=
(u-1)(u+2)\bar{S}_{1'2'}R^{ vs_+}_{31'}(u-\frac12)R^{
vs_-}_{32'}(u-\frac12)\bar{S}_{1'2'}^{-1}. \no\\\label{fu-12}\eea
In Eq.(\ref{fu-12}), we take the fusion in the auxiliary spaces
and obtain a 16-dimensional auxiliary space ${\rm\bf V}_{\langle
12 \rangle}$. We can show that the resulting  16-dimensional
auxiliary space is exact a direct tensor-product of two
4-dimensional auxiliary spaces $ {\rm\bf V^{(s_+)}}$ and $ {\rm\bf
V^{(s_-)}}$, which endows the (anti) spinor representation of
$so(6)$ respectively. The matrix $S_{1'2'}$ is defined in the
direct product space ${\rm\bf V^{(s_+)}}\otimes {\rm\bf
V^{(s_-)}}$ with the matrix form
\begingroup
\renewcommand*{\arraystretch}{0.1}
\begin{equation}
 S_{1'2'}=\begin{pmatrix}\setlength{\arraycolsep}{1.5pt}
 \begin{array}{cccc|cccc|cccc|cccc}
   1 &&& &&&& &&&& &&&& \\
    &&& &-1&&& &&&& &&&& \\
    &-1&& &&&& &&&& &&&& \\
    &&& &&-1&& &&&& &&&& \\
   \hline &&&b_2 &&&-b_2& &&-b_2&& &b_2&&& \\
   &&1& &&&& &&&& &&&& \\
    &&& &&&& &1&&& &&&& \\
    &&&b_2 &&&b_2& &&b_2&& &b_2&&& \\
   \hline &&& &&&& &&&1& &&&& \\
    &&&b_2 &&&b_2& &&-b_2&& &-b_2&&& \\
    &&& &&&& &&&& &&-1&& \\
    &&& &&&&  &&&& &&&-1& \\
   \hline &&&-a_2 &&&a_2& &&-a_2&& &a_2&&& \\
    &&& &&&&-1 &&&& &&&& \\
    &&& &&&& &&&&-1 &&&& \\
    &&& &&&& &&&& &&&& -1
\end{array}
    \end{pmatrix},\no
\end{equation}
\begin{equation}
 \bar{S}_{1'2'}=\begin{pmatrix}\setlength{\arraycolsep}{1.5pt}
 \begin{array}{cccc|cccc|cccc|cccc}
   1 &&& &&&& &&&& &&&& \\
    &&& &-1&&& &&&& &&&& \\
    &-1&& &&&& &&&& &&&& \\
    &&& &&-1&& &&&& &&&& \\
   \hline &&&b_2 &&&-b_2& &&-b_2&& &b_2&&& \\
   &&1& &&&& &&&& &&&& \\
    &&& &&&& &1&&& &&&& \\
    &&&b_2 &&&b_2& &&b_2&& &b_2&&& \\
   \hline &&& &&&& &&&1& &&&& \\
    &&&b_2 &&&b_2& &&-b_2&& &-b_2&&& \\
    &&& &&&& &&&& &&-1&& \\
    &&& &&&&  &&&& &&&-1& \\
   \hline &&&a_2 &&&-a_2& &&a_2&& &-a_2&&& \\
    &&& &&&&-1 &&&& &&&& \\
    &&& &&&& &&&&-1 &&&& \\
    &&& &&&& &&&& &&&& -1
\end{array}
    \end{pmatrix},\no
\end{equation}
\endgroup
 where \bea a_2=\frac{\sqrt{3}}{2}, \quad b_2=\frac12. \no
 \eea
 The relation between $ {S}_{1'2'}$ and $ \bar{S}_{1'2'}$ is
 provided by (\ref{ss}).
In Eq.(\ref{fu-12}), the matrices $R^{ s_{\pm}v}_{1'2}(u)$ are
defined in the ${\rm\bf V^{(s_{\pm})}}\otimes {\rm\bf V}$ space
and take the forms of
\begingroup
\renewcommand*{\arraystretch}{0.1}
\begin{equation}
R^{  s_+v}_{1'2}=
    \begin{pmatrix}\setlength{\arraycolsep}{1.0pt}
    \begin{array}{cccccc|cccccc|cccccc|cccccc}
    a_1&&&&& &&&&&& &&&&&& &&&&&&  \\
    &a_1&&&& &&&&&& &&&&&& &&&&&& \\
    &&a_1&&& &&&&&& &&&&&& &&&&&&  \\
    &&&b_1&& &&-1&&&& &-1&&&&& &&&&&&  \\
    &&&&b_1& &&&1&&& &&&&&& &-1&&&&&  \\
    &&&&&b_1 &&&&&& &&&1&&& &&1&&&& \\
   \hline &&&&& &a_1&&&&& &&&&&& &&&&&&  \\
    &&&-1&& &&b_1&&&& &-1&&&&& &&&&&& \\
    &&&&1& &&&b_1&&& &&&&&& &1&&&&& \\
    &&&&& &&&&a_1&& &&&&&& &&&&&&  \\
     &&&&& &&&&&a_1& &&&&&& &&&&&& \\
     &&&&& &&&&&&b_1 &&&&&1& &&&&-1&&  \\
   \hline  &&&-1&& &&-1&&&& &b_1&&&&& &&&&&&  \\
    &&&&& &&&&&& &&a_1&&&& &&&&&&  \\
   &&&&&1 &&&&&& &&&b_1&&& &&-1&&&&  \\
   &&&&& &&&&&& &&&&a_1&& &&&&&& \\
     &&&&& &&&&&&1 &&&&&b_1& &&&&1&&  \\
      &&&&& &&&&&& &&&&&&a_1 &&&&&&  \\
   \hline &&&&-1& &&&1&&& &&&&&& &b_1&&&&&  \\
    &&&&&1 &&&&&& &&&-1&&& &&b_1&&&&  \\
    &&&&&  &&&&&& &&&&&& &&&a_1&&& \\
     &&&&& &&&&&&-1 &&&&&1& &&&&b_1&& \\
     &&&&& &&&&&& &&&&&& &&&&&a_1&  \\
      &&&&& &&&&&& &&&&&& &&&&&&a_1  \\
\end{array}
    \end{pmatrix},\no
\end{equation}
\endgroup

\begingroup
\renewcommand*{\arraystretch}{0.1}
\begin{equation}
R^{  s_-v}_{1'2}=\begin{pmatrix}\setlength{\arraycolsep}{1.0pt}
    \begin{array}{cccccc|cccccc|cccccc|cccccc}
    a_1&&&&& &&&&&& &&&&&& &&&&&&  \\
    &a_1&&&& &&&&&& &&&&&& &&&&&& \\
    &&b_1&&& &&-1&&&& &-1&&&&& &&&&&&  \\
    &&&a_1&& &&&&&& &&&&&& &&&&&&  \\
    &&&&b_1& &&&&1&& &&&&&& &-1&&&&&  \\
    &&&&&b_1 &&&&&& &&&&1&& &&1&&&& \\
   \hline &&&&& &a_1&&&&& &&&&&& &&&&&&  \\
    &&-1&&& &&b_1&&&& &-1&&&&& &&&&&& \\
    &&&&& &&&a_1&&& &&&&&& &&&&&& \\
    &&&&1& &&&&b_1&& &&&&&& &1&&&&&  \\
     &&&&& &&&&&a_1& &&&&&& &&&&&& \\
     &&&&& &&&&&&b_1 &&&&&1& &&&-1&&&  \\
   \hline  &&-1&&& &&-1&&&& &b_1&&&&& &&&&&&  \\
    &&&&& &&&&&& &&a_1&&&& &&&&&&  \\
   &&&&& &&&&&& &&&a_1&&& &&&&&&  \\
   &&&&&1 &&&&&& &&&&b_1&& &&-1&&&& \\
     &&&&& &&&&&&1 &&&&&b_1& &&&1&&&  \\
      &&&&& &&&&&& &&&&&&a_1 &&&&&&  \\
   \hline &&&&-1& &&&&1&& &&&&&& &b_1&&&&&  \\
    &&&&&1 &&&&&& &&&&-1&& &&b_1&&&&  \\
    &&&&&  &&&&&&-1 &&&&&1& &&&b_1&&& \\
     &&&&& &&&&&& &&&&&& &&&&a_1&& \\
     &&&&& &&&&&& &&&&&& &&&&&a_1&  \\
      &&&&& &&&&&& &&&&&& &&&&&&a_1  \\
\end{array}
    \end{pmatrix},\no
\end{equation}
\endgroup
where the matrix elements are \bea a_1=u+\frac32, \quad
b_1=u+\frac12.\no \eea The matrices $R^{  s_{\pm}v}_{1'2}(u)$ have
the properties
\begin{eqnarray}
\hspace{-0.8truecm}{\rm unitarity}&:&R^{   s_{\pm}v}_{1'2}(u)R^{  vs_{\pm}}_{21'}(-u)=\rho_{s}(u)=a_1(u)a_1(-u),\nonumber\\[4pt]
\hspace{-0.8truecm}{\rm crossing \,\,unitarity}&:&R^{
s_{\pm}v}_{1'2}(u)^{t_{1'}}R^{
 vs_{\pm}}_{21'}(-u-4)^{t_{1'}}=\rho_{s}(u+2),\label{Properties}
\end{eqnarray}
and satisfy the Yang-Baxter equations
\bea
\hspace{-0.8truecm} R^{ s_{\pm}v}_{1'2}(u_1-u_2)R^{
s_{\pm}v}_{1'3}(u_1-u_3)R^{ vv}_{23}(u_2-u_3) =R^{
vv}_{23}(u_2-u_3)R^{ s_{\pm}v}_{1'3}(u_1-u_3)R^{
s_{\pm}v}_{1'2}(u_1-u_2).\label{QYB11-2}\eea

Now, we consider the degenerate case of matrix $R^{  s_{+}v}_{1'2}(u)$. At the point of $u=-\frac32$, we have \bea
R^{s_+v}_{1'2}(-\frac32)=P_{1'2}^{(+)}S_{1'2}^{(+)},\eea
where $S_{1'2}^{(+)}$ is a constant matrix omitted here and $P_{1'2}^{(+)}$ is a 4-dimensional projector
\bea
P_{1'2}^{(+)}=\sum_{i=1}^{4}
|{\phi}^{(+)}_i\rangle\langle{\phi}^{(+)}_i|,\no \eea
with the basis vectors
 \bea
&&|{\phi}^{(+)}_1\rangle=\frac{1}{\sqrt{3}}(|14\rangle+|22\rangle+|31\rangle),\quad
|{\phi}^{(+)}_2\rangle=\frac{1}{\sqrt{3}}(|15\rangle-|23\rangle+|41\rangle),\nonumber\\[4pt]
&&|{\phi}^{(+)}_3\rangle=\frac{1}{\sqrt{3}}(|16\rangle-|33\rangle-|42\rangle),\quad
|{\phi}^{(+)}_4\rangle=\frac{1}{\sqrt{3}}(|26\rangle-|35\rangle+|44\rangle).\no
\eea Taking the fusion by using the projector $P_{1'2}^{(+)}$, we obtain
\bea
&&P^{ (+) }_{1'2}R^{vv} _{23}(u)R^{s_+v}
_{1'3}(u-\frac32)P^{ (+) }_{1'2}=(u-1)(u+2)R^{s_-v}
_{\langle 1'2\rangle 3}(u-\frac12), \no \\
&&P^{ (+) }_{21'}R^{vv} _{32}(u)R^{vs_+} _{31'}(u-\frac32)P^{ (+)
}_{21'}=(u-1)(u+2)R^{vs_-} _{3\langle 1'2\rangle}(u-\frac12). \label{sv-2}
 \eea
We note that the dimension of the fused auxiliary space ${\rm\bf V}_{\langle 1'2\rangle}$ is 4.

At the point of $u=-\frac32$, we also have
\bea
R^{s_-v}_{1'2}(-\frac32)=P_{1'2}^{(-)}S_{1'2}^{(-)},\eea
where $P_{1'2}^{(-)}$ is a 4-dimensional projector
\bea
P_{1'2}^{(-)}=\sum_{i=1}^{4}
|{\phi}^{(-)}_i\rangle\langle{\phi}^{(-)}_i|, \no \eea
with the basis vectors
\bea
&&|{\phi}^{(-)}_1\rangle=\frac{1}{\sqrt{3}}(|13\rangle+|22\rangle+|31\rangle),\quad
|{\phi}^{(-)}_2\rangle=\frac{1}{\sqrt{3}}(|15\rangle-|24\rangle+|41\rangle),\nonumber\\[4pt]
&&|{\phi}^{(-)}_3\rangle=\frac{1}{\sqrt{3}}(|16\rangle-|34\rangle-|42\rangle),\quad
|{\phi}^{(-)}_4\rangle=\frac{1}{\sqrt{3}}(|26\rangle-|35\rangle+|43\rangle),\no
\eea
and the $S_{1'2}^{(-)}$ is a constant matrix omitted here. Taking the fusion by using the projector $P_{1'2}^{(-)}$,
we obtain \bea &&P^{ (-) }_{1'2}R^{vv} _{23}(u)R^{s_-v}
_{1'3}(u-\frac32)P^{ (-) }_{1'2}=(u-1)(u+2)R^{s_+v}
_{\langle 1'2\rangle 3}(u-\frac12), \no \\
 &&P^{ (-) }_{21'}R^{vv} _{32}(u)R^{vs_-}
_{31'}(u-\frac32)P^{ (-) }_{21'}=(u-1)(u+2)R^{vs_+}
_{3\langle 1'2\rangle }(u-\frac12). \label{sv-1111} \eea

\subsection{The operator product identities}

From the fused $R$-matrices $R^{s_{\pm}v}_{0'j}$, we can define the fused monodromy matrices
\bea
T_{0'}^{
\pm}(u)=R^{    s_{\pm}v}_{0'1}(u-\theta_1)R^{
s_{\pm}v}_{0'2}(u-\theta_2)\cdots R^{  s_{\pm}v}_{0'N}(u-\theta_N),\label{Transfer-S-}
\eea
which satisfy the Yang-Baxter relations
\bea
R^{  vs_{\pm}}_{00'}(u-v) T_0(u) T_{0'}^{\pm}(v) = T_{0'}^{\pm}(v)T_0(u)  R^{  vs_{\pm}}_{00'}(u-v). \label{y1bta2o}
\eea
Taking the partial trace in the auxiliary space, we obtain the fused transfer matrices
\bea
 t^{(p)}_{\pm}(u)=tr_{0'} T_{0'}^{\pm}(u).
\eea

From the Yang-Baxter relations (\ref{ybta2o}) and (\ref{y1bta2o}) at certain points and using the properties of projectors, we obtain
\bea &&T_1(\theta_j)\,T_2(\theta_j-{{2}})=
P^{{  vv}(1) }_{21}\,T_1(\theta_j)\,T_2(\theta_j-{{2}}), \no \\[6pt]
&&T_1(\theta_j)\,T_2(\theta_j-1)=P^{{  vv} (16) }_{21}\,T_1(\theta_j)\,T_2(\theta_j-1),\no \\[6pt]
&&T_2(\theta_j)\, T_{1'}^{ +}(\theta_j-\frac32)=
P_{1'2}^{(+)}\,T_2(\theta_j)\, T_{1'}^{ +}(\theta_j-\frac32),\no \\[6pt]
&&T_2(\theta_j)\, T_{1'}^{ -}(\theta_j-\frac32)=
P_{1'2}^{(-)}\,T_2(\theta_j)\, T_{1'}^{ -}(\theta_j-\frac32).
\label{opr-1} \eea
By using the fusion identities (\ref{hhgg-1}), (\ref{fu-12}), (\ref{sv-2}) and (\ref{sv-1111}), we have following fusion identities
\bea &&P^{{  vv}(1) }_{21}T_1(u)\,T_2(u-{{2}})P^{{  vv}(1) }_{21}=
P^{{  vv}(1) }_{21}\prod_{i=1}^N
a(u-\theta_i)e(u-\theta_i-{{2}})\times {\rm id}, \no \\[6pt]
&&P^{{  vv} (16) }_{21}T_1(u)\,T_2(u-1)P^{{  vv} (16) }_{21}=\prod_{i=1}^N
\tilde{\rho}_0(u-\theta_i)S_{1'2'}\,T^+_{1'}(u-\frac12)\,T^-_{2'}(u-\frac12)S_{1'2'}^{-1},\no \\[6pt]
&&P_{1'2}^{(+)}T_2(u)\, T_{1'}^{ +}(u-\frac32)P_{1'2}^{(+)}=
\prod_{i=1}^N
\tilde{\rho}_0(u-\theta_i)\, T_{\langle 1'2\rangle }^{ -}(u-\frac12),\no \\[6pt]
&&P_{1'2}^{(-)}T_2(u)\, T_{1'}^{ -}(u-\frac32)P_{1'2}^{(-)}=
\prod_{i=1}^N \tilde{\rho}_0(u-\theta_i)\, T_{\langle 1'2\rangle }^{
+}(u-\frac12),\label{opr-11}
 \eea
where \bea \tilde{\rho}_0(u)=(u-1)(u+2).\no \eea
Taking the partial trace in the auxiliary spaces and using the relations (\ref{opr-1}) and (\ref{opr-11}), we obtain following operator product identities
\bea && t^{(p)}(\theta_j)\,t^{(p)}(\theta_j-{{2}})=\prod_{i=1}^N
a(\theta_j-\theta_i)e(\theta_j-\theta_i-{{2}})\times {\rm id}, \no  \\
&& t^{(p)}(\theta_j)\,t^{(p)}(\theta_j-1)= \prod_{i=1}^N
\tilde{\rho}_0(\theta_j-\theta_i)\,t^{(p)}_+(\theta_j-\frac{1}{{2}})\,t^{(p)}_{-}(\theta_j-\frac{1}{{2}}),\no  \\
&&t^{(p)}(\theta_j)\,t^{(p)}_{+}(\theta_j-\frac{3}{2})=\prod_{i=1}^N
\tilde{\rho}_0(\theta_j-\theta_i)\,t^{(p)}_{-}(\theta_j-\frac{1}{2}),\no \\
&&t^{(p)}(\theta_j)\,t^{(p)}_{-}(\theta_j-\frac{3}{2})=\prod_{i=1}^N
\tilde{\rho}_0(\theta_j-\theta_i)\,t^{(p)}_{+}(\theta_j-\frac{1}{2}).\label{fut1p-5}\eea
From the definitions, the asymptotic behaviors of the transfer matrices can be calculated directly as
\bea && t^{ (p)}(u)|_{u\rightarrow \pm\infty}= 6u^{2N}\times {\rm  id} +\cdots,\no \\[4pt]
&& t^{(p)}_{\pm}(u)|_{u\rightarrow \pm\infty}=
4u^{N}\times {\rm   id} +\cdots.\label{fuwwwtpl-7} \eea

Let us denote the eigenvalues of the transfer matrices $t^{(p)}(u)$ and $t^{(p)}_{\pm}(u)$ as
$\Lambda^{(p)}(u)$ and $\Lambda^{(p)}_{\pm}(u)$, respectively.
We note that the eigenvalues $\Lambda^{(p)}(u)$ and $\Lambda^{(p)}_{\pm}(u)$ are the
polynomials of $u$ with degrees $2N$ and $N$, respectively. Thus we need $4N+3$ conditions to
determine the polynomials $\Lambda^{(p)}(u)$ and $\Lambda^{(p)}_{\pm}(u)$.

From the operator product identities (\ref{fut1p-5}), we have the functional relations
among the eigenvalues
\bea &&
\Lambda^{(p)}(\theta_j)\,\Lambda^{(p)}(\theta_j-{{2}})=\prod_{i=1}^N
a(\theta_j-\theta_i)e(\theta_j-\theta_i-{{2}}),\no  \\
&& \Lambda^{(p)}(\theta_j)\,\Lambda^{(p)}(\theta_j-1)=
\prod_{i=1}^N
\tilde{\rho}_0(\theta_j-\theta_i)\,\Lambda^{(p)}_+(\theta_j-\frac{1}{{2}})\,\Lambda^{(p)}_{-}(\theta_j-\frac{1}{{2}}),\no  \\
&&\Lambda^{(p)}(\theta_j)\,\Lambda^{(p)}_{+}(\theta_j-\frac{3}{2})=\prod_{i=1}^N
\tilde{\rho}_0(\theta_j-\theta_i)\,\Lambda^{(p)}_{-}(\theta_j-\frac{1}{2}),\no \\
&&\Lambda^{(p)}(\theta_j)\,\Lambda^{(p)}_{-}(\theta_j-\frac{3}{2})=\prod_{i=1}^N
\tilde{\rho}_0(\theta_j-\theta_i)\,\Lambda^{(p)}_{+}(\theta_j-\frac{1}{2}).\label{l1-4}\eea
And the corresponding asymptotic behaviors read
\bea && \Lambda^{ (p)}(u)|_{u\rightarrow \pm\infty}= 6u^{2N} +\cdots,\no \\[4pt]
&& \Lambda^{(p)}_{\pm}(u)|_{u\rightarrow \pm\infty}= 4u^{N}
+\cdots.\label{fuwww2tpl-7} \eea Then we arrive at that $4N$
functional relations (\ref{l1-4}) together with $3$ asymptotic
behaviors (\ref{fuwww2tpl-7}) give us sufficient conditions to
determine the eigenvalues.

\subsection{The $T-Q$ relations}

Let us introduce some  functions \bea
&&Z^{(p)}_1(u)=\prod_{j=1}^N
a(u-\theta_j)\,\frac{Q_{p}^{(1)}(u-1)}{Q_{p}^{(1)}(u)},\no\\[4pt]
&&Z^{(p)}_2(u)=\prod_{j=1}^Nb(u-\theta_j)\,\frac{Q_{p}^{(1)}(u+1)Q_{p}^{(2)}(u-1)Q_{p}^{(3)}(u-1)}{Q_{p}^{(1)}(u)Q_{p}^{(2)}(u)Q_{p}^{(3)}(u)},\no\\[4pt]
&&Z^{(p)}_3(u)=\prod_{j=1}^N
b(u-\theta_j)\,\frac{Q_{p}^{(2)}(u-1)Q_{p}^{(3)}(u+1)}{Q_{p}^{(2)}(u)Q_{p}^{(3)}(u)},\no\\[4pt]
&&Z^{(p)}_4(u)=\prod_{j=1}^N
b(u-\theta_j)\,\frac{Q_{p}^{(2)}(u+1)Q_{p}^{(3)}(u-1)}{Q_{p}^{(2)}(u)Q_{p}^{(3)}(u)},\no\\[4pt]
&&Z^{(p)}_5(u)=\prod_{j=1}^N
b(u-\theta_j)\,\frac{Q_{p}^{(1)}(u)Q_{p}^{(2)}(u+1)Q_{p}^{(3)}(u+1)}{Q_{p}^{(1)}(u+1)Q_{p}^{(2)}(u)Q_{p}^{(3)}(u)},\no\\
&&Z^{(p)}_6(u)=\prod_{j=1}^N e(u-\theta_j)\,
\frac{Q_{p}^{(1)}(u+2)}{Q_{p}^{(1)}(u+1)},\no\\[4pt]
&&Q_{p}^{(1)}(u)=\prod_{k=1}^{L_1}(u-\mu_k^{(1)}+\frac{1}{2}),\quad
Q_{p}^{(2)}(u)=\prod_{k=1}^{L_2}(u-\mu_k^{(2)}+1),\no
\\&& Q_{p}^{(3)}(u)=\prod_{k=1}^{L_3}(u-\mu_k^{(3)}+1).\eea The
relations (\ref{l1-4}) and (\ref{fuwww2tpl-7}) enable us to
parameterize the eigenvalues of the transfer matrices in terms of
some homogeneous  $T-Q$ relations as
\bea
&&\Lambda^{(p)}(u)=Z^{(p)}_1(u)+
Z^{(p)}_2(u)+Z^{(p)}_3(u)+Z^{(p)}_4(u)+Z^{(p)}_5(u)+Z^{(p)}_6(u), \no \\[4pt]
&&\Lambda^{(p)}_+(u)=\prod_{i=1}^N
a_1(u-\theta_i)\left[\frac{Q^{(2)}_p(u-\frac32)}{Q^{(2)}_p(u-\frac{1}{2})}+\frac{Q^{(1)}_p(u-\frac12)Q^{(2)}_p(u+\frac12)}{Q^{(1)}_p(u+\frac12)Q^{(2)}_p(u-\frac{1}{2})}\right]\no\\
&&\hspace{15mm}+\prod_{i=1}^N
b_1(u-\theta_i)\left[\frac{Q^{(3)}_p(u+\frac32)}{Q^{(3)}_p(u+\frac{1}{2})}+\frac{Q^{(1)}_p(u+\frac32)Q^{(3)}_p(u-\frac12)}{Q^{(1)}_p(u+\frac12)Q^{(3)}_p(u+\frac{1}{2})}\right],\no \\
&&\Lambda^{(p)}_-(u)=\prod_{i=1}^N
a_1(u-\theta_i)\left[\frac{Q^{(3)}_p(u-\frac32)}{Q^{(3)}_p(u-\frac{1}{2})}+\frac{Q^{(1)}_p(u-\frac12)Q^{(3)}_p(u+\frac12)}{Q^{(1)}_p(u+\frac12)Q^{(3)}_p(u-\frac{1}{2})}\right]\no\\
&&\hspace{15mm}+\prod_{i=1}^N
b_1(u-\theta_i)\left[\frac{Q^{(2)}_p(u+\frac32)}{Q^{(2)}_p(u+\frac{1}{2})}+\frac{Q^{(1)}_p(u+\frac32)Q^{(2)}_p(u-\frac12)}{Q^{(1)}_p(u+\frac12)Q^{(2)}_p(u+\frac{1}{2})}\right].
\label{T-Q-Periodic}
\eea The regularities of the eigenvalues $\Lambda^{(p)}(u)$ and $\Lambda_{\pm}^{(p)}(u)$ lead to that the Bethe roots $\{\mu^{(m)}_k\}$
should satisfy the Bethe ansatz equations (BAEs): \bea &&
\frac{Q_{p}^{(1)}(\mu_k^{(1)}-\frac{3}{2})Q_{p}^{(2)}(\mu_k^{(1)}-\frac{1}{2})Q_{p}^{(3)}(\mu_k^{(1)}-\frac{1}{2})}
{Q_{p}^{(1)}(\mu_k^{(1)}+\frac{1}{2})Q_{p}^{(2)}(\mu_k^{(1)}-\frac{3}{2})Q_{p}^{(3)}(\mu_k^{(1)}-\frac{3}{2})}
=-\prod_{j=1}^N \frac{\mu_k^{(1)}-\frac{1}{2}-\theta_j
}{\mu_k^{(1)}+\frac{1}{2}-\theta_j}, \no \\[6pt]&&
\qquad\qquad\qquad  k=1,\cdots, L_1, \no \\[6pt]
&&\frac{Q_{p}^{(1)}(\mu_l^{(2)})Q_{p}^{(2)}(\mu_l^{(2)}-{2})}{Q_{p}^{(1)}(\mu_l^{(2)}-1)Q_{p}^{(2)}(\mu_l^{(2)})}
=-1, \quad l=1,\cdots, L_2.\no  \\[6pt]
&&\frac{Q_{p}^{(1)}(\mu_l^{(3)})Q_{p}^{(3)}(\mu_l^{(3)}-{2})}{Q_{p}^{(1)}(\mu_l^{(3)}-1)Q_{p}^{(3)}(\mu_l^{(3)})}
=-1, \quad l=1,\cdots, L_3.\label{BAEs-3} \eea
We note that the BAEs obtained from the regularity of
$\Lambda^{(p)}(u)$ are the same as those obtained from the regularity
of $\Lambda^{(p)}_{\pm}(u)$. It is easy to check that $\Lambda^{(p)} (u)$
and $\Lambda_{\pm}^{(p)}(u)$ satisfy the functional relations
(\ref{l1-4}) and the asymptotic behaviors
(\ref{fuwww2tpl-7}). Therefore, we conclude
that $\Lambda^{(p)}(u)$ and $\Lambda_{\pm}^{(p)}(u)$ are the eigenvalues of
the transfer matrices $t^{(p)}(u)$ and $t^{(p)}_{\pm}(u)$, respectively.  It is easy to check that the homogeneous $T-Q$ relations  (\ref{T-Q-Periodic}) and associated
BAEs (\ref{BAEs-3}) coincide with those obtained previously by others methods \cite{NYReshetikhin1, NYReshetikhin2, Bn}.

Then the eigenvalues of the Hamiltonian (\ref{haha-1}) can be obtained by the $\Lambda^{(p)}(u)$ as
\begin{eqnarray}
E_p= \frac{\partial \ln \Lambda^{(p)}(u)}{\partial
u}|_{u=0,\{\theta_j\}=0}.
\end{eqnarray}

\section{$D^{(1)}_3$ model with non-diagonal open boundary condition}
\setcounter{equation}{0}

In this section, we consider the system with general integrable open boundary condition.
Let us introduce a pair of reflection matrices $K^{v}(u)$ and $\bar K^{v}(u)$. The former satisfies the reflection equation (RE)
\begin{equation}
 R^{   vv}_{12}(u-v){K^{  v}_{  1}}(u)R^{   vv}_{21}(u+v) {K^{   v}_{2}}(v)=
 {K^{   v}_{2}}(v)R^{   vv}_{12}(u+v){K^{   v}_{1}}(u)R^{   vv}_{21}(u-v),
 \label{r1}
 \end{equation}
and the latter  satisfies the dual RE
\begin{eqnarray}
 &&R^{   vv}_{12}(-u+v){\bar{K}^{   v}_{1}}(u)R^{  vv}_{21}
 (-u-v-4){\bar{K}^{   v}_{2}}(v)\nonumber\\[4pt]
&&\qquad\qquad\quad\quad={\bar{K}^{   v}_{2}}(v)R^{
vv}_{12}(-u-v-4) {{\bar{K}}^{ v}_{1}}(u)R^{  vv}_{21}(-u+v).
 \label{r2}
 \end{eqnarray}
For the open case, instead of the
``row-to-row" monodromy matrix $T^{v}_0(u)$ given by (\ref{Mon-1}), one needs
consider  the ``double-row" monodromy matrix to describe the reflection process. Let us introduce another ``row-to-row" monodromy matrix
\begin{eqnarray}
\hat{T}_0^{  v} (u)=R_{N0}^{  vv}(u+\theta_N)\cdots R_{20}^{  vv}(u+\theta_{2}) R_{10}^{  vv}(u+\theta_1),\label{Tt11}
\end{eqnarray}
which satisfies the Yang-Baxter relation
\begin{eqnarray}
R_{ 12}^{  vv} (u-v) \hat T_{1}^{  v}(u) \hat T_2^{  v}(v)=\hat  T_2^{  v}(v) \hat T_{ 1}^{  v}(u) R_{12}^{  vv} (u-v).\label{haishi0}
\end{eqnarray}
The transfer matrix $t(u)$ is defined as
\begin{equation}
t(u)= tr_0 \{\bar K_0^{  v }(u)T_0^{  v} (u) K^{  v }_0(u)\hat{T}^{  v}_0 (u)\}. \label{trweweu1110}
\end{equation}
From the Yang-Baxter relation, reflection equation and its dual, one can
prove that the transfer matrices with different spectral parameters
commute with each other, $[t(u), t(v)]=0$. Therefore, $t(u)$ serves
as the generating function of all the conserved quantities of the
system.  The Hamiltonian is constructed by taking the
derivative of the logarithm of the transfer matrix
\begin{eqnarray}
H&=&\frac{\partial \ln t(u)}{\partial
u}|_{u=0,\{\theta_j\}=0} \nonumber \\[8pt]
&=& \sum^{N-1}_{k=1}H_{k k+1}+\frac{{K^{v}_1}(0)'}{2\xi}+\frac{
tr_0 \{\bar K^{v}_0(0)H_{N0}\}}{tr_0 \bar K^{v}_0(0)}+{\rm
constant}, \label{hh}
\end{eqnarray} where $H_{k\,k+1}$ is given by (\ref{local-Ham}) and $ K^{v}(0)=\xi\times {\rm   id}.$

\subsection{Reflection matrix}

   In this paper, we consider the general
integrable open boundary condition where the reflection matrix has
the non-diagonal elements which breaks the $U(1)$-symmetry of the
system. The general reflection matrix for  $D_n^{(1)}$ vertex
model with its R-matrix being trigonometric has been obtained by
Lima-Santos and Malara \cite{lima-1,lima-2}. Here we chose the
reflection matrix $K^{v-}(u)$, which is the solution of reflection
equation (\ref{r1}), as \bea
K^{v}(u)=\left(\begin{array}{cccccc}K_{11}^{v}(u)&0&0&0&0&0\\[6pt]
    0&K_{22}^{v}(u)&0&K_{24}^{v}(u)&0&0\\[6pt]
    0&0&K_{33}^{v}(u)&0&K_{35}^{v}(u)&0\\[6pt]
    0&K_{42}^{v}(u)&0&K_{44}^{v}(u)&0&0\\[6pt]
    0&0&K_{53}^{v}(u)&0&K_{55}^{v}(u)&0\\[6pt]
0&0&0&0&0&K_{66}^{v}(u)\end{array}\right),\label{K-matrix-VV}\eea
where the non-vanishing matrix elements are
\bea
&&K_{11}^{v}(u)=\frac{(2-c_2-2c_2u)(2+c_2+2c_2u)}{2}, \no\\
&& K_{22}^{v}(u)=\frac{(2-c_2-2cu)(2+c_2+2c_2u)}{2},\no\\
&&K_{24}^{v}(u)=-c_1u(2+c_2+2c_2u), \no\\
&& K_{33}^{v}(u)=\frac{(2+c_2-2cu)(2-c_2-2c_2u)}{2},
\no\\
&&K_{35}^{v}(u)=c_1u(2-c_2-2c_2u) ,\no\\
&& K_{42}^{v}(u)=-c_3u(2+c_2+2c_2u),\no\\
&& K_{44}^{v}(u)=\frac{(2-c_2+2cu)(2+c_2+2c_2u)}{2},
\no\\
&& K_{53}^{v}(u)=-c_3u(-2+c_2+2c_2u),\no\\
 && K_{55}^{v}(u)=\frac{(2+c_2+2cu)(2-c_2-2c_2u)}{2},
 \no\\
 &&K_{66}^{v}(u)=\frac{(2-c_2-2c_2u)(2+c_2+2c_2u)}{2}.
 \label{K-matrix-3}\eea
Here $c_1$, $c_2$ and $c_3 $ are  free boundary parameters, where
$c$ is expressed in terms of them as \bea c^2=c_2^2-c_1c_3. \no
\eea The Eq.(\ref{K-matrix-VV}) can also be obtained at some
special free parameters by taking rational limit to the general
reflection matrix in \cite{lima-1,lima-2}.

The dual reflection matrix $\bar K^{v}(u)$ is also a non-diagonal
one and given by
\begin{equation}
\bar K^{   v}(u)=K^{
v}(-u-2)|_{(c_1, c_2, c_3)\rightarrow\,
(c'_1, c'_2, c'_3)}, \label{ksk111}
\end{equation}
where  $c'_1$, $c'_2$ and $c'_3$ are the boundary parameters at the other side. For a generic choice of the three boundary parameters $\{c_i,\,c'_i|i=1,2,3\}$,
it is easily to check that
$[K^{v}(u),\,\bar K^{v}(v)]\neq 0$. This implies that the matrices
$K^{v}(u)$ and $\bar K^{v}(u)$ cannot be diagonalized simultaneously.

Following the method developed in \cite{Yan-06} and using the
crossing-symmetry (\ref{Crossing-symmetry}) of the $R$-matrix, the explicit expressions (\ref{K-matrix-VV}) and (\ref{ksk111}) of the
reflection matrices, we find that the transfer matrix (\ref{trweweu1110}) possesses the crossing symmetry
\begin{eqnarray}
t(-u-2)=t(u).\label{Transfer-Crossing}
\end{eqnarray}

\subsection{Fusion of the reflection matrix}

In order to seek the relations that the transfer matrix (\ref{trweweu1110}) satisfies, we first consider the fusion of the reflection matrix. The fusion of the one-dimensional projector $P_{12}^{ vv(1)}$ gives
\bea && P_{12}^{ vv(1)}K_{2}^{ v}(u)R_{12}^{ vv}(2u-2)K_{2}^{ v}(u-2)P_{21}^{ vv(1)}\no\\
&&\qquad  =(u-2)(u-\frac{3}{{2}})h(u)h(-u)P_{12}^{ vv(1)},\no \\
&& P_{21}^{ vv(1)}\bar{K}_{1}^{ v}(u-2)R_{21}^{ vv}(-2u-2)\bar{K}_{2}^{ v}(u)P_{12}^{\rm vv(1)}\no\\
&& \qquad =(u+2)(u+\frac{3}{{2}})\tilde{h}(u)\tilde{h}(-u)P_{21}^{
    vv(1)},  \label{0805-1} \eea where \bea
h(u)=(2-c_2-2c_2u)(2+c_2+2c_2u),\quad
\tilde{h}(u)=(2-c'_2-2c'_2u)(2+c'_2+2c'_2u). \eea

The fusion of the 16-dimensional projector $P_{12}^{ vv(16)}$ gives
\bea && P_{21}^{ vv(16)}K^{
    v}_1(u)R^{ vv}_{21}(2u-1)K^{ v}_2(u-1)P_{12}^{ vv(16)}\no\\
    &&\qquad =2(u-1)h(u)S_{1'2'}K^{
    s_+}_{1'}(u-\frac{1}{2})R^{ s_-s_+}_{2'1'}(2u-1)K^{
    s_-}_{2'}(u-\frac{1}{2})\bar{S}_{1'2'}^{-1}, \no \\
&&P^{ vv(16)}_{12}\bar{K}^{
    v}_2(u-1)R^{ vv}_{12}(-2u-3)\bar{K}^{v}_1(u)P^{ vv(16)}_{21}\no\\&&\qquad  =-2(u+2)\tilde{h}(u)\bar{S}_{1'2'}\bar{K}^{
    s_-}_{1'}(u-\frac{1}{2})R^{ s_+s_-}_{1'2'}(-2u-3)\bar{K}^{
    s_+}_{2'}(u-\frac{1}{2})S_{1'2'}^{-1},  \label{0805-2} \eea
where
\bea
 &&R^{  s_+s_-}_{1'2'}=\left(\begin{array}{cccc|cccc|cccc|cccc}
    a_3&&& &&&& &&&& &&&& \\
    &a_3&& &&&& &&&& &&&& \\
    &&a_3& &&&& &&&& &&&& \\
    &&&b_3 &&&1& &&-1&& &1&&& \\
   \hline &&& &a_3&&& &&&& &&&& \\
   &&& &&a_3&& &&&& &&&& \\
    &&&1 &&&b_3& &&1&& &-1&&& \\
    &&& &&&&a_3 &&&& &&&& \\
   \hline &&& &&&& &a_3&&& &&&& \\
    &&&-1 &&&1& &&b_3&& &1&&& \\
    &&& &&&& &&&a_3& &&&& \\
    &&& &&&&  &&&&a_3 &&&& \\
   \hline &&&1 &&&-1& &&1&& &b_3&&& \\
    &&& &&&& &&&& &&a_3&& \\
    &&& &&&& &&&& &&&a_3& \\
    &&& &&&& &&&& &&&& a_3\\
           \end{array}\right), \no
\eea
\bea
&&K^{s_+}(u)=\left(\begin{array}{cccc}1-c u&c_1u&0&0\\[6pt]
    c_3u&1+cu&0&0\\[6pt]
    0&0&1+c_2u&0\\[6pt]
    0&0&0&1-c_2u\end{array}\right),\label{K-matrix-1} \\
&&K^{s_-}(u)=\left(\begin{array}{cccc}1+c_2 u&0&0&0\\[6pt]
0&1-c_2u&0&0\\[6pt]
    0&0&1-cu&c_1u\\[6pt]
    0&0&    c_3u&1+cu\end{array}\right),\label{K-matrix-2} \eea
and \bea a_3=u+2, \quad b_3=u+1. \no \eea
The matrix $R^{  s_+s_-}_{1'2'}(u)$ has properties
\begin{eqnarray}
{\rm unitarity}&:&R^{   s_+s_-}_{1'2'}(u)R^{  s_-s_+}_{2'1'}(-u)=(2+u)(2-u),\nonumber\\[4pt]
{\rm crossing\,\,unitarity}&:&R^{ s_+s_-}_{1'2'}(u)^{t_{1'}}R^{
 s_-s_+}_{2'1'}(-u-4)^{t_{1'}}=\rho_{ss}(u)=-(u+1)(u+3),
\end{eqnarray}
and satisfies the Yang-Baxter equations
\bea
&&R^{
s_+s_-}_{1'2'}(u_1-u_2)R^{ s_+v}_{1'3}(u_1-u_3)R^{
s_-v}_{2'3}(u_2-u_3)\no \\
&& =R^{ s_-v}_{2'3}(u_2-u_3)R^{
s_+v}_{1'3}(u_1-u_3)R^{ s_+s_-}_{1'2'}(u_1-u_2). \eea
The matrices $K^{s_{\pm}}(u)$ satisfy the reflection equations
\bea
&& R^{   s_{\pm}v}_{12}(u-v){K^{  s_{\pm}}_{  1}}(u)R^{   vs_{\pm}}_{21}(u+v) {K^{   v}_{2}}(v) \no \\
&&\qquad =
 {K^{   v}_{2}}(v)R^{   s_{\pm}v}_{12}(u+v){K^{   s_{\pm}}_{1}}(u)R^{   vs_{\pm}}_{21}(u-v).
 \label{1r2}
 \eea
 The $\bar{S}_{1'2'}$ can be obtained by \bea\bar{S}_{1'2'}=\frac 12{S}_{1'2'}R^{
s_-s_+}_{2'1'}(0).\label{ss}\eea The dual reflection matrices
$\bar{K}^{s_{\pm}}(u)$ are constructed as  \bea
\bar{K}^{s_{\pm}}(u)=K^{s_{\pm}}(-u-2)\left|_{(c,c_1,c_2,c_3)\rightarrow
   (c',c_1',c_2',c_3')}\right., \eea
which satisfy the dual REs
\begin{eqnarray}
 &&R^{   s_{\pm}v}_{12}(-u+v){\bar{K}^{   s_{\pm}}_{1}}(u)R^{  vs_{\pm}}_{21}
 (-u-v-4){\bar{K}^{   v}_{2}}(v)\nonumber\\[4pt]
&&\qquad\qquad\quad\quad={\bar{K}^{   v}_{2}}(v)R^{
s_{\pm}v}_{12}(-u-v-4) {{\bar{K}}^{s_{\pm}}_{1}}(u)R^{  vs_{\pm}}_{21}(-u+v).
 \label{1dr3}
 \end{eqnarray}

The fusion of the 4-dimensional projector $P_{1'2}^{(\pm)}$ gives
\bea && P_{1'2}^{(\pm)}K_2^{v}(u)R_{1'2}^{ s_{\pm}v}(2u-\frac{3}{2})K_{1'}^{ s_{\pm}}(u-\frac{3}{2})P_{21'}^{({\pm})} =(u-\frac{3}{2})h(u)K_{\langle 1'2\rangle }^{ s_{\mp}}(u-\frac{1}{2}),\no \\
&& P_{21'}^{(\pm)}\bar{K}_{1'}^{ s_{\pm}}(u-\frac{3}{2})R_{21'}^{ vs_{\pm}}(-2u-\frac{5}{2})\bar{K}_2^{v}(u)P_{1'2}^{(\pm)} =
-(u+2)\tilde{h}(u)\bar{K}_{\langle 1'2\rangle }^{ s_{\mp}}(u-\frac{1}{2}). \label{0805-3} \eea

\subsection{Operators product relations}

Define the fused monodromy matrices $\hat{T}_{ {0}'}^{\pm}(u)$ by using the fused $R$-matrix $R^{    vs_{\pm}}$ as
\begin{eqnarray}
\hat{T}_{ {0}'}^{ \pm}(u)=R^{    vs_{\pm}}_{N {0'}}(u+\theta_N)\cdots R^{
vs_{\pm}}_{2 {0'}}(u+\theta_{2}) R^{
   vs_{\pm}}_{1 {0'}}(u+\theta_1),\label{M2on-2}
\end{eqnarray}
which satisfy the Yang-Baxter relation
\begin{eqnarray}
R_{00'}^{ vs_{\pm}} (u-v) \hat T_{0}(u) \hat T_{0'}^{  \pm}(v)=\hat  T_{0'}^{  \pm}(v) \hat T_{0}(u) R_{00'}^{ vs_{\pm}} (u-v).\label{haish1i0}
\end{eqnarray}
From the Yang-Baxter relations (\ref{haishi0}) and (\ref{haish1i0}) at certain points and using the properties of projectors, we obtain
\bea &&\hat{T}_1(-\theta_j)\,\hat{T}_2(-\theta_j-{{2}})=
P^{{  vv}(1) }_{12}\,\hat{T}_1(-\theta_j)\,\hat{T}_2(-\theta_j-{{2}}), \no \\[6pt]
&&\hat{T}_1(-\theta_j)\,\hat{T}_2(-\theta_j-1)=P_{12}^{{  vv}(16) }\,\hat{T}_1(-\theta_j)\,\hat{T}_2(-\theta_j-1),\no \\[6pt]
&&\hat{T}_2(-\theta_j)\, \hat{T}_{1'}^{ +}(-\theta_j-\frac32)=
P_{21'}^{(+)}\,\hat{T}_2(-\theta_j)\, \hat{T}_{1'}^{ +}(-\theta_j-\frac32),\no \\[6pt]
&&\hat{T}_2(-\theta_j)\, \hat{T}_{1'}^{ -}(-\theta_j-\frac32)=
P_{21'}^{(-)}\,\hat{T}_2(-\theta_j)\, \hat{T}_{1'}^{
-}(-\theta_j-\frac32). \label{op22r-1} \eea
By using the fusion identities (\ref{hhgg-1}), (\ref{fu-12}), (\ref{sv-2}) and (\ref{sv-1111}), we also obtain the fusion identities
\bea &&P^{{  vv}(1) }_{12}\hat{T}_1(u)\,\hat{T}_2(u-{{2}})P^{{
vv}(1) }_{12}= P^{{  vv}(1) }_{12}\prod_{i=1}^N
a(u+\theta_i)e(u+\theta_i-{{2}})\times {\rm id}, \no \\[6pt]
&&P^{{  vv}(16) }_{12}\hat{T}_1(u)\,\hat{T}_2(u-1)P^{{  vv}(16) }_{12}=\prod_{i=1}^N
\tilde{\rho}_0(u+\theta_i)S_{1'2'}\,\hat{T}^+_{1'}(u-\frac12)\,\hat{T}^-_{2'}(u-\frac12)S_{1'2'}^{-1},\no \\[6pt]
&&P_{21'}^{(+)}\hat{T}_2(u)\, \hat{T}_{1'}^{
+}(u-\frac32)P_{21'}^{(+)}= \prod_{i=1}^N
\tilde{\rho}_0(u+\theta_i)\, \hat{T}_{\langle 1'2\rangle }^{ -}(u-\frac12),\no \\[6pt]
&&P_{21'}^{(-)}\hat{T}_2(u)\, \hat{T}_{1'}^{
-}(u-\frac32)P_{21'}^{(-)}= \prod_{i=1}^N
\tilde{\rho}_0(u+\theta_i)\, \hat{T}_{\langle 1'2\rangle }^{
+}(u-\frac12).\label{opr-1012}
 \eea

The fused transfer matrices are defined as
\bea
t_{\pm}(u)=tr_{0'}\{\bar{K}^{s_{\pm}}_{0'}(u) T_{0'}^{\pm}(u)K^{
s_{\pm}}_{0'}(u)\hat{T}_{0'}^{\pm}(u)\}. \eea
Direct calculation shows
\bea &&
t(u)t(u+\Delta)=[\rho_1(2u+\Delta+2)]^{-1}tr_{12}\{\bar{K}^{v}_{2}(u+\Delta)R_{12}^{vv}(-2u-4-\Delta) \no\\
&&\hspace{10mm}\times \bar{K}_1^{ v}(u)T_1 (u) T_{2}(u+\Delta)K^{
v}_1(u)R_{21}^{vv}(2u+\Delta)K^{ v}_{2}(u+\Delta)\hat{T}_1
(u)\hat{T}_{2}(u+\Delta)\},\label{tt-1}  \\
&&t(u)t_{\pm}(u+\Delta)=[\rho_s(2u+\Delta+2)]^{-1}tr_{12}\{\bar{K}^{s_{\pm}}_{2}(u+\Delta)R_{12}^{vs_{\pm}}(-2u-4-\Delta) \no\\
&&\hspace{5mm}\times \bar{K}_1^{ v}(u)T_1 (u)
T_{2}^{\pm}(u+\Delta)K^{ v}_1(u)R_{21}^{s_{\pm}v}(2u+\Delta)K^{
s_{\pm}}_{2}(u+\Delta)\hat{T}_1
(u)\hat{T}_{2}^{{\pm}}(u+\Delta)\}, \label{tt-2} \\
 &&
t_+(u)t_-(u+\Delta)=[\rho_{ss}(2u+\Delta)]^{-1}tr_{12}\{\bar{K}^{s_{-}}_{2}(u+\Delta)R_{12}^{s_+s_{-}}(-2u-4-\Delta) \no\\
&&\hspace{10mm}\times \bar{K}_1^{ s_+}(u)T_1^+ (u)
T_{2}^{-}(u+\Delta)K^{ s_+}_1(u)\no \\
&& \hspace{10mm}\times R_{21}^{s_{-}s_+}(2u+\Delta)K^{
s_{-}}_{2}(u+\Delta)\hat{T}_1^+
(u)\hat{T}_{2}^{-}(u+\Delta)\}.\label{0804-12} \eea Here $\Delta$
choose $-2,-1,-\frac32,0$ for
Eq.(\ref{tt-1}),Eq.(\ref{tt-1}),Eq.(\ref{tt-2}) and
Eq.(\ref{0804-12}), respectively.

With the help of relations (\ref{opr-1}), (\ref{opr-11}),
(\ref{0805-1}), (\ref{0805-2}), (\ref{0805-3}), (\ref{op22r-1}),
(\ref{opr-1012}) and considering Eqs.(\ref{tt-1}-\ref{0804-12}) at
certain points, we arrive at \bea &&
t(\pm\theta_j)\,t(\pm\theta_j-{{2}})=\frac{1}{2^4}\frac{(\pm\theta_j-2)(\pm\theta_j+2)(\pm\theta_j-\frac32)(\pm\theta_j+\frac32)}
{(\pm\theta_j-1)(\pm\theta_j+1)(\pm\theta_j-\frac12)(\pm\theta_j+\frac12)}h(\pm\theta_j)h(\mp\theta_j)\tilde{h}(\pm\theta_j)\no\\
&&\quad \times\tilde{h}(\mp\theta_j)\prod_{i=1}^N
a(\pm\theta_j-\theta_i)e(\pm\theta_j-\theta_i-{{2}})a(\pm\theta_j+\theta_i)e(\pm\theta_j+\theta_i-{{2}})\times {\rm id},\no  \\
&&
t(\pm\theta_j)\,t(\pm\theta_j-1)=\frac{(\pm\theta_j-1)(\pm\theta_j+2)}{(\pm\theta_j-\frac12)(\pm\theta_j+\frac32)}\prod_{i=1}^N
\tilde{\rho}_0(\pm\theta_j-\theta_i)\tilde{\rho}_0(\pm\theta_j+\theta_i)\no\\
&&\hspace{25mm}\times h(\pm\theta_j)\tilde{h}(\pm\theta_j) t_+(\pm\theta_j-\frac{1}{{2}})\,t_{-}(\pm\theta_j-\frac{1}{{2}}),\no \\
&&t(\pm\theta_j)\,t_{+}(\pm\theta_j-\frac{3}{2})=\frac{1}{2^2}\frac{(\pm\theta_j-\frac32)(\pm\theta_j+2)}{(\pm\theta_j-\frac12)(\pm\theta_j+1)}\prod_{i=1}^N
\tilde{\rho}_0(\pm\theta_j-\theta_i)\tilde{\rho}_0(\pm\theta_j+\theta_i)\no\\
&&\hspace{25mm}\times h(\pm\theta_j)\tilde{h}(\pm\theta_j) t_{-}(\pm\theta_j-\frac{1}{2}),\no \\
&&t(\pm\theta_j)\,t_{-}(\pm\theta_j-\frac{3}{2})=\frac{1}{2^2}\frac{(\pm\theta_j-\frac32)(\pm\theta_j+2)}{(\pm\theta_j-\frac12)(\pm\theta_j+1)}\prod_{i=1}^N
\tilde{\rho}_0(\pm\theta_j-\theta_i)\tilde{\rho}_0(\pm\theta_j+\theta_i)\no\\
&&\hspace{25mm}\times h(\pm\theta_j)\tilde{h}(\pm\theta_j)
t_{+}(\pm\theta_j-\frac{1}{2}).\label{Op-Product-Periodic-41} \eea
Meanwhile, from the definitions we also know the values of transfer matrices at some special points
\bea &&
t(0)=\frac32(2-c_2)(2+c_2)(2-c'_2)(2+c'_2)\prod_{l=1}^N\rho_1
(-\theta_l), \no\\
&& t(-\frac12)=6\prod_{l=1}^N\rho_s (-\theta_l)t_{\pm}(-1), \no\\
&&t_+(0)=t_-(0)=4\prod_{l=1}^N\rho_s
(-\theta_l). \label{2Op} \eea
The corresponding asymptotic behaviors read
\bea
&&t(u)|_{u\rightarrow \pm\infty}=-\frac{8c_2c'_2((c'_3)^2c^2+c'^2c_3^2-c_3^2(c'_2)^2-c_2^2(c'_3)^2-2cc'c_3c'_3-c_2c_3c'_2c'_3)}{c_3c'_3}u^{4N+4}\times {\rm id}\no \\
&&\qquad \qquad\qquad +\cdots, \no \\
&&t_+(u)|_{u\rightarrow \pm\infty}=\frac{(c'c_3-c'_3c-c'_2c_3-c'_3c_2)(c'c_3-c'_3c+c'_2c_3+c'_3c_2)}{c_3c'_3}u^{2N+2} \times {\rm id}+\cdots, \no\\
&&t_-(u)|_{u\rightarrow \pm\infty}=\frac{(c'c_3-c'_3c-c'_2c_3-c'_3c_2)(c'c_3-c'_3c+c'_2c_3+c'_3c_2)}{c_3c'_3}u^{2N+2}\times {\rm id}+\cdots. \label{1Op}
\eea

Denote the eigenvalues of the fused transfer matrices $t(u)$ and $t_{\pm}(u)$ as
$\Lambda(u)$ and $\Lambda_{\pm}(u)$, respectively. From Eq.(\ref{Op-Product-Periodic-41} ), we obtain the functional relations among the eigenvalues of the transfer matrices
\bea &&
\Lambda(\pm\theta_j)\,\Lambda(\pm\theta_j-{{2}})=\frac{1}{2^4}\frac{(\pm\theta_j-2)(\pm\theta_j+2)(\pm\theta_j-\frac32)(\pm\theta_j+\frac32)}
{(\pm\theta_j-1)(\pm\theta_j+1)(\pm\theta_j-\frac12)(\pm\theta_j+\frac12)}h(\pm\theta_j)h(\mp\theta_j)\tilde{h}(\pm\theta_j)\no\\
&&\hspace{10mm}\times\tilde{h}(\mp\theta_j)\prod_{i=1}^N
a(\pm\theta_j-\theta_i)e(\pm\theta_j-\theta_i-{{2}})a(\pm\theta_j+\theta_i)e(\pm\theta_j+\theta_i-{{2}}),\no  \\
&&
\Lambda(\pm\theta_j)\,\Lambda(\pm\theta_j-1)=\frac{(\pm\theta_j-1)(\pm\theta_j+2)}{(\pm\theta_j-\frac12)(\pm\theta_j+\frac32)}\prod_{i=1}^N
\tilde{\rho}_0(\pm\theta_j-\theta_i)\tilde{\rho}_0(\pm\theta_j+\theta_i)\no\\
&&\hspace{25mm}\times h(\pm\theta_j)\tilde{h}(\pm\theta_j) \Lambda_+(\pm\theta_j-\frac{1}{{2}})\,\Lambda_{-}(\pm\theta_j-\frac{1}{{2}}),\no \\
&&\Lambda(\pm\theta_j)\,\Lambda_{+}(\pm\theta_j-\frac{3}{2})=\frac{1}{2^2}\frac{(\pm\theta_j-\frac32)(\pm\theta_j+2)}{(\pm\theta_j-\frac12)(\pm\theta_j+1)}\prod_{i=1}^N
\tilde{\rho}_0(\pm\theta_j-\theta_i)\tilde{\rho}_0(\pm\theta_j+\theta_i)\no\\
&&\hspace{25mm}\times h(\pm\theta_j)\tilde{h}(\pm\theta_j) \Lambda_{-}(\pm\theta_j-\frac{1}{2}),\no  \\
&&\Lambda(\pm\theta_j)\,\Lambda_{-}(\pm\theta_j-\frac{3}{2})=\frac{1}{2^2}\frac{(\pm\theta_j-\frac32)(\pm\theta_j+2)}{(\pm\theta_j-\frac12)(\pm\theta_j+1)}\prod_{i=1}^N
\tilde{\rho}_0(\pm\theta_j-\theta_i)\tilde{\rho}_0(\pm\theta_j+\theta_i)\no\\
&&\hspace{25mm}\times h(\pm\theta_j)\tilde{h}(\pm\theta_j)
\Lambda_{+}(\pm\theta_j-\frac{1}{2}).\label{Op-e4} \eea
Eqs.(\ref{2Op}) and (\ref{1Op}) give rise to the relations
\bea &&
\Lambda(0)=\frac32(2-c_2)(2+c_2)(2-c'_2)(2+c'_2)\prod_{l=1}^N\rho_1
(-\theta_l), \no\\
&& \Lambda(-\frac12)=6\prod_{l=1}^N\rho_s
(-\theta_l)\Lambda_{\pm}(-1), \no\\
&&\Lambda_+(0)=\Lambda_-(0)=4\prod_{l=1}^N\rho_s
(-\theta_l),\label{3Op}
\eea
and
\bea
&&\Lambda(u)|_{u\rightarrow \pm\infty}= -\frac{8c_2c'_2((c'_3)^2c^2+c'^2c_3^2-c_3^2(c'_2)^2-c_2^2(c'_3)^2-2cc'c_3c'_3-c_2c_3c'_2c'_3)}{c_3c'_3}u^{4N+4}
\no \\
&&\qquad\qquad \qquad+\cdots,\no\\
&&\Lambda_+(u)|_{u\rightarrow \pm\infty}=\frac{(c'c_3-c'_3c-c'_2c_3-c'_3c_2)(c'c_3-c'_3c+c'_2c_3+c'_3c_2)}{c_3c'_3}u^{2N+2}+\cdots,\no\\
&&\Lambda_-(u)|_{u\rightarrow
\pm\infty}=\frac{(c'c_3-c'_3c-c'_2c_3-c'_3c_2)(c'c_3-c'_3c+c'_2c_3+c'_3c_2)}{c_3c'_3}u^{2N+2}+\cdots.\label{4Op}
\eea From the definitions, we know that the eigenvalues
$\Lambda(u)$ and $\Lambda_{\pm}(u)$ are the polynomials of $u$
with degrees $4N+4$ and $2N+2$, respectively. Meanwhile,
$\Lambda(u)$ and $\Lambda_{\pm}(u)$ enjoy the crossing symmetries
\bea \Lambda(-u-{2})=\Lambda(u), \quad
\Lambda_{\pm}(-u-{2})=\Lambda_{\mp}(u). \eea Therefore, in order
to determine the explicit expressions of the polynomials, we need
$4N+7$ conditions, which are all listed by
Eqs.(\ref{Op-e4})-(\ref{4Op}).

\subsection{Inhomogeneous T-Q relations}

Let us introduce some  functions \bea
&&Z_1(u)=2^2\frac{(u+2)(u+\frac32)}{(u+1)(u+\frac12)}\prod_{j=1}^N
a(u-\theta_j)a(u+\theta_j)\no\\
&&\hspace{15mm} \times h_1(u+\frac12)h_1(u-\frac12)\tilde{h}_1(u+\frac12)\tilde{h}_1(u-\frac12)\,\frac{Q^{(1)}(u-1)}{Q^{(1)}(u)},\no\\[4pt]
&&Z_2(u)=2^2\frac{u(u+2)(u+\frac32)}{(u+1)(u+1)(u+\frac12)}\prod_{j=1}^Nb(u-\theta_j)b(u+\theta_j)\no\\
&&\hspace{15mm} \times h_1(u+\frac12)h_2(u+\frac32)\tilde{h}_1(u+\frac12)\tilde{h}_2(u+\frac32)\frac{Q^{(1)}(u+1)Q^{(2)}(u-1)Q^{(3)}(u-1)}{Q^{(1)}(u)Q^{(2)}(u)Q^{(3)}(u)},\no\\[4pt]
&&Z_3(u)=2^2\frac{u(u+2)}{(u+1)(u+1)}\prod_{j=1}^N
b(u-\theta_j)b(u+\theta_j)\no\\
&&\hspace{15mm} \times h_1(u+\frac12)h_2(u+\frac32)\tilde{h}_1(u+\frac12)\tilde{h}_2(u+\frac32)\frac{Q^{(2)}(u-1)Q^{(3)}(u+1)}{Q^{(2)}(u)Q^{(3)}(u)},\no\\[4pt]
&&Z_4(u)=2^2\frac{u(u+2)}{(u+1)(u+1)}\prod_{j=1}^N
b(u-\theta_j)b(u+\theta_j)\no\\
&&\hspace{15mm} \times h_1(u+\frac12)h_2(u+\frac32)\tilde{h}_1(u+\frac12)\tilde{h}_2(u+\frac32)\frac{Q^{(2)}(u+1)Q^{(3)}(u-1)}{Q^{(2)}(u)Q^{(3)}(u)},\no\\[4pt]
&&Z_5(u)=2^2\frac{u(u+2)(u+\frac12)}{(u+1)(u+1)(u+\frac32)}\prod_{j=1}^Nb(u-\theta_j)b(u+\theta_j)\no\\
&&\hspace{15mm} \times h_1(u+\frac12)h_2(u+\frac32)\tilde{h}_1(u+\frac12)\tilde{h}_2(u+\frac32)\frac{Q^{(1)}(u)Q^{(2)}(u+1)Q^{(3)}(u+1)}{Q^{(1)}(u+1)Q^{(2)}(u)Q^{(3)}(u)},\no\\
&&Z_6(u)=2^2\frac{u(u+\frac12)}{(u+1)(u+\frac32)}\prod_{j=1}^N
e(u-\theta_j)e(u+\theta_j)\no\\
&&\hspace{15mm} \times
h_2(u+\frac52)h_2(u+\frac32)\tilde{h}_2(u+\frac52)\tilde{h}_2(u+\frac32)
\frac{Q^{(1)}(u+2)}{Q^{(1)}(u+1)},\no\\[4pt]
&&f_1(u)=2^2x\,
\frac{u(u+2)(u+\frac32)}{u+1}\prod_{j=1}^Na(u-\theta_j)a(u+\theta_j)(u-\theta_j)(u+\theta_j)\no\\
&&\hspace{15mm} \times h_1(u+\frac12)\tilde{h}_1(u+\frac12)\frac{Q^{(2)}(u-1)Q^{(3)}(u-1)}{Q^{(1)}(u)},\no\\
&&f_2(u)=2^2x\,
\frac{u(u+2)(u+\frac12)}{u+1}\prod_{j=1}^Na(u-\theta_j)a(u+\theta_j)(u-\theta_j)(u+\theta_j)\no\\
&&\hspace{15mm} \times
h_2(u+\frac32)\tilde{h}_2(u+\frac32)\frac{Q^{(2)}(u+1)Q^{(3)}(u+1)}{Q^{(1)}(u+1)}, \no
\eea
where
\bea
&&Q^{(1)}(u)=\prod_{k=1}^{L_1}(u-\mu_k^{(1)}+\frac{1}{2})(u+\mu_k^{(1)}+\frac{1}{2}),\no\\
&&Q^{(2)}(u)=\prod_{k=1}^{L_2}(u-\mu_k^{(2)}+1)(u+\mu_k^{(2)}+1),\no\\
&&Q^{(3)}(u)=\prod_{k=1}^{L_3}(u-\mu_k^{(3)}+1)(u+\mu_k^{(3)}+1), \no \\
&&h_1(u)=1+c_2u,\quad h_2(u)=1-c_2u,\quad
\tilde{h}_1(u)=1-c'_2u,\quad \tilde{h}_2(u)=1+c'_2u. \no
\eea
The constraints (\ref{Op-e4})-(\ref{4Op}) enables us to parameterize the eigenvalues of the transfer matrices $\Lambda(u)$ and $\Lambda_{\pm}(u)$ in terms of
the inhomogeneous $T-Q$ relations \bea &&\Lambda(u)=Z_1(u)+
Z_2(u)+Z_3(u)+Z_4(u)+Z_5(u)+Z_6(u)+f_1(u)+f_2(u), \no \\[4pt]
&&\Lambda_+(u)=\prod_{i=1}^N
a_1(u-\theta_i)a_1(u+\theta_i)h_1(u)\tilde{h}_1(u)\no\\
&&\hspace{15mm}\times \left[\frac{u+2}{u+\frac12}\frac{Q^{(2)}(u-\frac32)}{Q^{(2)}(u-\frac{1}{2})}
+\frac{u(u+2)}{(u+1)(u+\frac12)}\frac{Q^{(1)}(u-\frac12)Q^{(2)}(u+\frac12)}{Q^{(1)}(u+\frac12)Q^{(2)}(u-\frac{1}{2})}\right]\no\\
&&\hspace{15mm}+\prod_{i=1}^N
b_1(u-\theta_i)b_1(u+\theta_i)h_2(u+2)\tilde{h}_2(u+2)\no\\
&&\hspace{15mm}\times\left[\frac{u}{u+\frac32}\frac{Q^{(3)}(u+\frac32)}{Q^{(3)}(u+\frac{1}{2})}
+\frac{u(u+2)}{(u+1)(u+\frac32)}\frac{Q^{(1)}(u+\frac32)Q^{(3)}(u-\frac12)}{Q^{(1)}(u+\frac12)Q^{(3)}(u+\frac{1}{2})}\right]\no\\
&&\hspace{10mm}+x\,
u(u+2)\prod_{i=1}^Na_1(u-\theta_i)a_1(u+\theta_i)
b_1(u-\theta_i)b_1(u+\theta_i)\frac{Q^{(2)}(u+\frac{1}{2})Q^{(3)}(u-\frac{1}{2})}{Q^{(1)}(u+\frac12)}, \no\\
 &&\Lambda_-(u)=\prod_{i=1}^N
a_1(u-\theta_i)a_1(u+\theta_i)h_1(u)\tilde{h}_1(u)\no\\
&&\hspace{15mm}\times \left[\frac{u+2}{u+\frac12}\frac{Q^{(3)}(u-\frac32)}{Q^{(3)}(u-\frac{1}{2})}
+\frac{u(u+2)}{(u+1)(u+\frac12)}\frac{Q^{(1)}(u-\frac12)Q^{(3)}(u+\frac12)}{Q^{(1)}(u+\frac12)Q^{(3)}(u-\frac{1}{2})}\right]\no\\
&&\hspace{15mm}+\prod_{i=1}^N
b_1(u-\theta_i)b_1(u+\theta_i)h_2(u+2)\tilde{h}_2(u+2)\no\\
&&\hspace{15mm}\times\left[\frac{u}{u+\frac32}\frac{Q^{(2)}(u+\frac32)}{Q^{(2)}(u+\frac{1}{2})}
+\frac{u(u+2)}{(u+1)(u+\frac32)}\frac{Q^{(1)}(u+\frac32)Q^{(2)}(u-\frac12)}{Q^{(1)}(u+\frac12)Q^{(2)}(u+\frac{1}{2})}\right]\no\\
&&\hspace{10mm}+x\,
u(u+2)\prod_{i=1}^Na_1(u-\theta_i)a_1(u+\theta_i)
b_1(u-\theta_i)b_1(u+\theta_i)\frac{Q^{(2)}(u-\frac{1}{2})Q^{(3)}(u+\frac{1}{2})}{Q^{(1)}(u+\frac12)}. \no
\eea The regularities of the eigenvalues $\Lambda(u)$ and $\Lambda_{\pm}(u)$ lead to that the Bethe roots
$\{\mu^{(m)}_k\}$ should satisfy the BAEs \bea
&&\frac{(\mu_k^{(1)}+\frac12)h_1(\mu_k^{(1)}-1)\tilde{h}_1(\mu_k^{(1)}-1)}{\prod_{j=1}^N(\mu_k^{(1)}-\frac{1}{2}-\theta_j)(\mu_k^{(1)}-\frac{1}{2}+\theta_j)}
\frac{Q^{(1)}(\mu_k^{(1)}-\frac{3}{2})}{Q^{(2)}(\mu_k^{(1)}-\frac{3}{2})Q^{(3)}(\mu_k^{(1)}-\frac{3}{2})}\no\\
&&\hspace{5mm}
+\frac{(\mu_k^{(1)}-\frac12)h_2(\mu_k^{(1)}+1)\tilde{h}_2(\mu_k^{(1)}+1)}{\prod_{j=1}^N(\mu_k^{(1)}+\frac{1}{2}-\theta_j)(\mu_k^{(1)}+\frac{1}{2}+\theta_j)}
\frac{Q^{(1)}(\mu_k^{(1)}+\frac{1}{2})}{Q^{(2)}(\mu_k^{(1)}-\frac{1}{2})Q^{(3)}(\mu_k^{(1)}-\frac{1}{2})}\no\\
&&
=-x\,\mu_k^{(1)}(\mu_k^{(1)}-\frac12)(\mu_k^{(1)}+\frac12), \quad k=1,\cdots, L_1, \no \\[6pt]
&&\frac{Q^{(1)}(\mu_l^{(2)})Q^{(2)}(\mu_l^{(2)}-{2})}{Q^{(1)}(\mu_l^{(2)}-1)Q^{(2)}(\mu_l^{(2)})}
=-\frac{\mu_l^{(2)}-\frac12}{\mu_l^{(2)}+\frac12}, \quad l=1,\cdots, L_2, \no  \\[6pt]
&&\frac{Q^{(1)}(\mu_l^{(3)})Q^{(3)}(\mu_l^{(3)}-{2})}{Q^{(1)}(\mu_l^{(3)}-1)Q^{(3)}(\mu_l^{(3)})}
=-\frac{\mu_l^{(3)}-\frac12}{\mu_l^{(3)}+\frac12}, \quad
l=1,\cdots, L_3, \label{BAEs-223} \eea where the numbers of Bethe
roots should satisfy the constraint \bea L_1=L_2+L_3+N, \eea and
the parameter $x$ is given by \bea
x=\frac{(c'c_3-c'_3c-c'_2c_3-c'_3c_2)(c'c_3-c'_3c+c'_2c_3+c'_3c_2)}{c_3c'_3}+4
c_2c'_2.\eea Again, the BAEs obtained from the regularity of
$\Lambda(u)$ are the same as those obtained from the regularity of
$\Lambda_{\pm}(u)$. The function $Q^{(m)}(u)$ has two zero points,
and the BAEs obtained from these two points are the same. It is
easy to check  that $\Lambda(u)$ and $\Lambda_{\pm}(u)$ satisfy
the functional relations (\ref{Op-e4}), the values at the special
points (\ref{3Op}) and the asymptotic behaviors (\ref{4Op}).
Therefore, we conclude that $\Lambda(u)$ and $\Lambda_{\pm}(u)$
are the eigenvalues of the transfer matrices $t(u)$ and
$t_{\pm}(u)$, respectively.

Finally, the eigenvalue $E$ of Hamiltonian (\ref{hh}) can be obtained by the $\Lambda(u)$ as
\begin{eqnarray}
E= \frac{\partial \ln \Lambda(u)}{\partial
u}|_{u=0,\{\theta_j\}=0}.
\end{eqnarray}

\section{Discussion}

In this paper, we study the exact solution of $D^{(1)}_3$ model, with various  boundary conditions including the periodic one and the non-diagonal reflection one.
By using the fusion technique, we obtain the complete operator product identities of the fused transfer matrices.
Based on them and the asymptotic
behaviors as well as the special values at certain points, we obtain the Bethe Ansatz solutions of the system.
The method and the results in this paper could be generalized to the $D^{(1)}_n$ case directly.

\section*{Acknowledgments}

The financial supports from National Program for Basic Research of
MOST (Grant Nos. 2016 YFA0300600 and 2016YFA0302104), National
Natural Science Foundation of China (Grant Nos. 11934015,
11975183, 11547045, 11774397, 11775178, 11775177 and 91536115),
Major Basic Research Program of Natural Science of Shaanxi
Province (Grant Nos. 2017KCT-12, 2017ZDJC-32), Australian Research
Council (Grant No. DP 190101529), Strategic Priority Research
Program of the Chinese Academy of Sciences, and Double First-Class
University Construction Project of Northwest University are
gratefully acknowledged. J Cao, Y Wang and WL Yang would like to
thank Prof. Yao-Zhong Zhang at the school of Mathematics and
Physics of The University of Queensland for his hospitality, where
some part work of the paper has been done.  GL Li acknowledges the
support from Shaanxi Province Key Laboratory of Quantum
Information and Quantum Optoelectronic Devices, Xi'an Jiaotong
University.

\section*{Appendix A: Spinorial R-matrix}
\setcounter{equation}{0}
\renewcommand{\theequation}{A.\arabic{equation}}

In this Appendix, we show that the vectorial $R$-matrix $R^{ vv}(u)$ (\ref{0804-1}) and the fused ones $R^{s_{\pm}v}(u)$  (see above (\ref{Properties})) can be obtained from the spinorial $R$-matrix of
$D^{(1)}_3$ model \cite{spm} by using the fusion.
The spinorial $R$-matrix $R^{  ss}_{1'2'}(u)$ of the $D^{(1)}_3$ model is the fundamental $R$-matrix of $su(4)$ one, and is a $16\times 16$ matrix with the form \bea
R^{  ss}_{1'2'}(u)= \left(\begin{array}{cccc|cccc|cccc|cccc}
    u+1&&& &&&& &&&& &&&& \\
    &u&& &1&&& &&&& &&&& \\
    &&u& &&&& &1&&& &&&& \\
    &&&u &&&& &&&& &1&&& \\
   \hline &1&& &u&&& &&&& &&&& \\
   &&& &&u+1&& &&&& &&&& \\
    &&& &&&u& &&1&& &&&& \\
    &&& &&&&u &&&& &&1&& \\
   \hline &&1& &&&& &u&&& &&&& \\
    &&& &&&1& &&u&& &&&& \\
    &&& &&&& &&&u+1& &&&& \\
    &&& &&&&  &&&&u &&&1& \\
   \hline &&&1 &&&& &&&& &u&&& \\
    &&& &&&&1 &&&& &&u&& \\
    &&& &&&& &&&&1 &&&u& \\
    &&& &&&& &&&& &&&& u+1\\
           \end{array}\right).\nonumber
\eea
The spinorial $R$-matrix has following properties
\begin{eqnarray}
\hspace{-0.8truecm}{\rm unitarity}&:&R^{   ss}_{1'2'}(u)R^{  ss}_{2'1'}(-u)=(1+u)(1-u),\nonumber\\[4pt]
\hspace{-0.8truecm}{\rm crossing-unitarity}&:&R^{
ss}_{1'2'}(u)^{t_{1'}}R^{
 ss}_{2'1'}(-u-4)^{t_{1'}}=-u(u+4),
\end{eqnarray}
and satisfies the Yang-Baxter equation \bea
&&R^{
ss}_{1'2'}(u_1-u_2)R^{ ss}_{1'3'}(u_1-u_3)R^{
ss}_{2'3'}(u_2-u_3) \nonumber \\
&&=R^{ ss}_{2'3'}(u_2-u_3)R^{
ss}_{1'3'}(u_1-u_3)R^{
ss}_{1'2'}(u_1-u_2).\label{QYB1-2s}\eea

At the point of $u=-1$, we have \bea
R^{ss}_{1'2'}(-1)=P_{1'2'}^{(6)}S_{1'2'}^{(6)},\eea
where $P_{1'2'}^{(6)}$ is a 6-dimensional projector \bea
P_{1'2'}^{(6)}=\sum_{i=1}^{6}
|{\phi}^{(6)}_i\rangle\langle{\phi}^{(6)}_i|,\label{vs1} \eea
and the corresponding basis vectors are
 \bea
&&|{\phi}^{(6)}_1\rangle=\frac{1}{\sqrt{2}}(|12\rangle-|21\rangle),\quad
|{\phi}^{(6)}_2\rangle=\frac{1}{\sqrt{2}}(|13\rangle-|31\rangle),\quad
|{\phi}^{(6)}_3\rangle=\frac{1}{\sqrt{2}}(|14\rangle-|41\rangle),\nonumber\\[4pt]
&&|{\phi}^{(6)}_4\rangle=\frac{1}{\sqrt{2}}(|23\rangle-|32\rangle),\quad
|{\phi}^{(6)}_5\rangle=\frac{1}{\sqrt{2}}(|24\rangle-|42\rangle),\quad
|{\phi}^{(6)}_6\rangle=\frac{1}{\sqrt{2}}(|34\rangle-|43\rangle).\nonumber
\eea The $S_{1'2'}^{(6)}$ is a $6\times 6$ constant matrix omitted here.
The fusion of the 6-dimensional projector $P_{1'2'}^{(6)}$ gives
\bea &&P^{ (6) }_{2'3'}R^{ss}
_{1'2'}(u+\frac12)R^{ss} _{1'3'}(u-\frac12)P^{ (6)
}_{2'3'}=(u-\frac12)R^{s_+v}
_{1'\langle 2'3'\rangle }(u), \label{sv1+1}\\[4pt]
&&P^{ (6) }_{1'2'}R^{s_+v} _{2'3}(u+\frac12)R^{s_+v}
_{1'3}(u-\frac12)P^{ (6) }_{1'2'}=R^{vv} _{\langle 1'2'\rangle 3}(u).
\label{sv1+2}
 \eea
The dimension of the fused space ${\bf V}_{\langle 1'2'\rangle} $ is 6.
From Eq.(\ref{sv1+1}), we obtain the fused $R$-matrix $R^{ s_{+}v}(u)$. The $R^{ s_{-}v}(u)$ can be obtained via Eq.(\ref{sv-2}).
From Eq.(\ref{sv1+2}), we obtain the vectorial $R$-matrix $R^{ vv}(u)$ (\ref{0804-1}).

For the open case, the spinorial $R$-matrix $R^{  ss}(u)$ and the spinorial reflection matrix $K^{s}(u)$ satisfy the reflection equation \bea &&
R^{ss}_{1'2'}(u-v){K^{s}_{ 1'}}(u)R^{ ss}_{2'1'}(u+v)
{K^{ s}_{2'}}(v) \no\\
&&=
 {K^{  s}_{2'}}(v)R^{ ss}_{1'2'}(u+v){K^{ s}_{1'}}(u)R^{
 ss}_{2'1'}(u-v).
 \label{0R1}\eea
One can check that matrix (\ref{K-matrix-1}) is a solution of Eq.(\ref{0R1}), $K^{s_{+}}(u)=K^{s}(u)$. By using Eq.(\ref{0805-3}), we arrive at $K^{s_-}(u)$ (\ref{K-matrix-2}).
The vectorial reflection matrix
$K^{v}(u)$ (\ref{K-matrix-3}) is obtained from $K^{s_{+}}(u)$ by using the fusion with 6-dimensional projector $P_{1'2'}^{(6)}$
\bea P_{1'2'}^{(6)}K_{2'}^{ s_+}(u+\frac{1}{2})R_{1'2'}^{ ss}(2u)K_{1'}^{ s_+}(u-\frac{1}{2})P_{2'1'}^{(6)}=(u-\frac{1}{2})K_{\langle 1'2'\rangle}^{ v}(u).\eea


\end{document}